\documentclass[aps,prl,showpacs,twocolumn,byrevtex,floatfix]{revtex4-1}
\usepackage{amsmath,amssymb,natbib}
\usepackage{graphicx}
\usepackage{epsfig}
\usepackage{color}

\begin{document}

\title{Stranski-Krastanov mechanism of growth and the effect of misfit sign on
quantum dots nucleation}

\author{J.E. Prieto$^{1}$ and I. Markov$^{2*}$}

\affiliation{$^1$Centro de Microan\'alisis de Materiales,
Dpto. de F\'\i{}sica de la Materia Condensada, IFIMAC and Instituto
Universitario ``Nicol\'as Cabrera", Universidad Aut\'onoma de Madrid,
28049 Madrid, Spain\\
$^2$Institute of Physical Chemistry, Bulgarian Academy of Sciences, 1113
Sofia, Bulgaria}

\email{joseemilio.prieto@uam.es}
\email{imarkov@ipc.bas.bg}

\date{\today}
\begin{abstract}
The thermodynamics of the Stranski-Krastanov mode of epitaxial growth and the
effect of the sign of the lattice misfit are discussed. The
Stranski-Krastanov mode of growth represents a sequence of layer-by-layer or
Frank-van der Merwe growth followed by the formation of three-dimensional (3D)
islands or Volmer-Weber growth. The occurrence of both growth modes mentioned
above is in compliance with the wettability criterion of Bauer. The positive
wetting function required for the occurrence of the Volmer-Weber growth is
originated by the vertical displacements of the atoms close to the edges of the
two-dimensional (2D) islands as a result of the relaxation of the lattice
misfit. The monolayer high islands become unstable against bilayer islands,
bilayer islands in turn become unstable against trilayer islands, etc. beyond
some critical islands sizes. Monolayer islands appear as necessary precursors of
three-dimensional (3D) islands. The critical island size for mono-bilayer
transformation increases steeply with decreasing lattice misfit and diverges at
a critical value of the misfit. This value divides the regions of Frank-van der
Merwe and Stranski-Krastanov modes in a phase diagram of coordinates
wetting-misfit. The transformation of monolayer to multilayer islands takes
place either by consecutive nucleation and growth of 2D islands (layer-by-layer
transformation), or by nucleation and lateral (2D) growth of multilayer islands
(multilayer 2D transformation). 
The former occurs in the case of ``stiff" overlayer
materials and mostly in compressed overlayers. The latter takes place in the
case of ``soft" materials like Pb and In, mostly in tensile overlayers. Tensile
films show non-nucleation transformation compared with the nucleation-like
behavior of compressed films.

\end{abstract}


\maketitle

\section{Introduction}

In 1958 Ernst Bauer published his famous thermodynamic criterion for the
classification of the mechanisms of epitaxial growth.\cite{Bauer58,Kern79} He
derived an expression for the equilibrium shape, given by the ratio $h/l$ (height/width),
of a cubic crystal on a foreign substrate in terms of the interrelation of the
specific surface energies of the substrate $\sigma_s$, epilayer, $\sigma$, and
the substrate-epilayer interface, $\sigma_i$. The change of the surface energy,
$\Delta \sigma = \sigma + \sigma_i - \sigma_s$ associated with the
formation of the epilayer, represents in fact a measure of the wetting of
the substrate
by the film material. In the case of incomplete wetting, $\Delta \sigma > 0$,
the growth proceeds by the formation and growth of separate
three-dimensional (3D) islands, a mechanism for which Bauer coined the
term Volmer-Weber (VW) growth.\cite{Volmer26}
When $\Delta \sigma \le 0$ and the lattice misfit is negligible, the height of
the 3D island is equal to zero and two-dimensional (2D) islands form
instead giving rise to layer-by-layer or Frank-van der Merwe (FM)
growth.\cite{Frank49,Frank491} And finally, when $\Delta \sigma < 0$ and the
lattice misfit is non-zero the growth begins by the formation of
a {\it wetting layer} consisting of a few monolayers-thick film followed by
the growth of 3D islands on top.  This is the well-known
Stranski-Krastanov mode of growth.\cite{Stranski39}

The equilibrium shape of a crystal on an unlike substrate had been earlier
derived by Kaischew in 1950 in terms of the binding energies between two
atoms of the deposit (cohesion energy, $\psi$) and between an atom of the 
substrate and an atom of the film (adhesion energy, 
$\psi^{prime}$).\cite{Kaischew50,Kaischew60} Both expressions, due to Bauer 
and Kaischew, respectively, for the equilibrium shape are in fact 
identical.\cite{Markov17} The condition
$\psi^{\prime} < \psi$ is equivalent to $\Delta \sigma > 0$, $\psi^{\prime} =
\psi$ corresponds to $\Delta \sigma = 0$ and $\psi^{\prime} > \psi$ corresponds
to $\Delta \sigma < 0$. It follows that the mechanism of growth depends
on the interrelation of the cohesion and adhesion energies. As will be shown
below, the lattice misfit plays a crucial role only when 
$\psi^{\prime} \ge \psi$
($\Delta \sigma \le 0$). Note that the above conclusions about the mechanism of
growth are based on the concept of the equilibrium crystal shape.

As shown by Rudolf Peierls, the mechanism of growth is closely connected
with the sign of the derivative of the chemical potential with respect to
the number of atoms in the overlayer, $d\mu /dN$.\cite{Peierls78} As seen in
Fig.~\ref{tdf}, the VW growth is associated with $d\mu /dN < 0$ and the FM
growth requires the condition $d\mu /dN > 0$. This means that the VW growth is
connected with a negative curvature, $d^2G/dN^2 < 0$, of the $N$-dependence of
the Gibbs free energy of the thickening film, whereas the FM growth is connected
with the opposite behavior, $d^2G/dN^2 > 0$. This implies that in the case
of SK growth, the dependence of the Gibbs free energy on film thickness must
possess an inflection point, $N_{\rm i}$, at which the curvature of $G$,
$d^2G/dN^2$,  changes sign from positive to negative with increasing film
thickness. The analysis of the problem shows that the planar film is stable up
to some critical thickness, $N_{\rm cr}$, which is slightly smaller
than $N_{\rm i}$. At $N = N_{\rm cr}$ $\mu = \mu_{\infty}$ and $P = P_{\infty}$
where $\mu_{\infty}$ and $P_{\infty}$ are the chemical potential and the
equilibrium vapor pressure of the infinitely large bulk deposit crystal,
respectively. Thus $N_{\rm cr}$ and $N_{\rm i}$ determine the thicknesses of the
stable and unstable wetting layers, which are given in Fig.\ \ref{tdf} by 
the lower dotted and the upper straight lines, respectively. Note that in the
analysis of Peierls the dependences of the film Gibbs free energies on film
thickness are smooth and differentiable, which results in $\Delta \mu (N_{\rm
cr}) = 0$. The analysis of Peierls leads to the same criterion as the one
derived by Bauer $\Delta \sigma = \sigma + \sigma_i - \sigma_s \gtrless 0$. For
more details the reader is referred to section 4.3.4 of Ref.
(\onlinecite{Markov17}).

It is obvious that the SK growth represents an instability of the planar growth
against clustering owing to the accumulation of strain energy in the wetting
layer. This led to the concept of nucleation of islands
due to the trade-off between the cost of the additional surface energy of the
3D islands and the gain of energy due to the elastic relaxation of the 3D
islands relative to the wetting layer\cite{Tersoff93,Tersoff94,Politi00}.
Although this approach gives a valuable insight into the problem, it does not
allow the determination of the mechanism of formation of the 3D islands on top
of the wetting layer. The essence of the problem is that the coherent
(dislocationless) SK mode consists of the formation of 3D islands of a material
$A$ on the same (strained) material $A$.\cite{Eaglesham90}

On the other hand, Mo {\it et al.}\cite{Mo90} observed with the help of
scanning tunneling microscopy (STM) Ge islands representing elongated 
pyramids (``hut"
clusters) bounded by (105) facets. The authors suggested that the hut clusters
are a step in the pathway to the formation of larger islands with steeper side
walls.\cite{Sutter04,Lutz94,Brehm11} The ways of relaxation of lattice misfit
in the transition from hut clusters to larger islands with steeper side
facets has been reviewed by Teichert.\cite{Teichert02} Tersoff {\it et al.}
have shown that the growth of SiGe superlattices up to 2000 layers resulted in a
very narrow size distribution of the quantum dots.\cite{Tersoff96}. Chen {\it et
al.}\cite{Chen97} and Vailionis {\it et al.}
\cite{Vailionis00} studied the initial stages of formation of the hut clusters
and found three- to four monolayers-high prepyramids with rounded bases in a
narrow interval of Ge coverages. Sutter and Lagally~\cite{Sutter00} suggested
another scenario for the formation SiGe alloy clusters at low misfit. They
observed by low-energy electron microscopy (LEEM) the formation of an array
of stepped mounds (ripples) as precursors of the hut clusters. These ripples
are inherent to strained films to relax the misfit strain as suggested by many
authors.\cite{Asaro72,Grinfeld86,Srolovitz89,Pimpinelli98} Based on these
observations, Sutter and Lagally suggested the concept of barrierless
(nucleationless) formation of the 3D islands.\cite{Sutter00} Similar views on
the idea of barrierless transformation of the ripples into faceted islands were
suggested by Tromp {\it et al.}\cite{Tromp00} and by Tersoff {\it et
al.}\cite{Tersoff02} The contradiction of the above-mentioned concepts of
nucleation and nucleationless formation of 3D islands, as well as many other
aspects of the growth modes gave rise to intensive theoretical studies of the
Stranski-Krastanov morphology by making use of both analytical
approaches\cite{Ratsch93,Merwe00,Kukushkin02}, and computer Monte
Carlo\cite{Petrov12,Biehl03,Khor00,Avery97} and molecular
dynamics\cite{Ashu91,Roland93,Joyce97,Xu04} simulations, and were debated in
numerous review papers and
monographs.\cite{Dubrovskii14,Shchukin99,Shchukin03,Bhatta15}. However, among
the most important questions remains the following: Is the nucleation concept of
3D clustering consistent with the wettability concept of Bauer?

In addition, the mechanism of growth of quantum dots in the SK mode depends
strongly on the sign of the lattice misfit. In compressed overlayers the film
atoms interact through the steeper repulsive branch of the interatomic
potential, whereas in tensile overlayers the interaction through the
weaker attractive branch prevails. The anharmonicity of the chemical bonding
influences the adhesion of the 3D islands to the wetting layer or, in other
words, the wettability as defined by Bauer, through the relaxation of strain
both laterally (in-plane) and vertically (out-of-plane) at the steps
forming the boundaries of the islands. This strain relaxation leads to two
different mechanisms of 2D-3D transformation, the consecutive transformations of
islands with gradually increasing height by nucleation of single monolayers,
and a mechanism in which multilayer islands nucleate and then laterally 
(two-dimensionally) grow. 
Note that the two-dimensional multilayer islands grow only laterally
keeping their height constant in contrast to three-dimensional islands which
grow both in length and height.

The paper is organized as follows. In consecutive sections we consider the
equilibrium vapor pressure of the 2D and 3D phases, the effect of lattice misfit
on the film-substrate adhesion, the thickness of the stable wetting layer, the
stability of mono- and multilayer islands, the layer-by-layer growth of 3D
islands, and the multilayer growth of 3D islands. We then compare our findings
with experimental data and discuss the results.

\section{Equilibrium vapor pressure of the 2D and 3D phases}

In 1929 Stranski\cite{Stranski129,Stranski229} studied the stability of 
separate monolayers of a monovalent ionic crystal K$^+$A$^-$ on the surface 
of the isomorphous divalent crystal K$^{2+}$A$^{2-}$ by making use of 
the newly discovered concept of the half-crystal or kink 
position.\cite{Kossel27,Stranski27,Stranski28} He found that the 
equilibrium vapor pressure of the first momolayer, $P_1$, is much
lower than the equilibrium vapor pressure, $P_0$, of the bulk monovalent
crystal. The reason is that the monovalent ions are attracted by the underlying
divalent ions more strongly than by the corresponding monovalent ions of the
same crystal. As the ions of the second monolayer are repulsed more strongly by
the underlying divalent ions of the substrate crystal, its equilibrium vapor
pressure will be higher than the equilibrium pressure $P_0$. The equilibrium
vapor pressure of the third monolayer will be smaller than $P_0$, and that of
the fourth monolayer will be already nearly equal to $P_0$, i.e. the 
energetic influence of the divalent substrate disappears beyond 
four monolayers. Thus they concluded
that the chemical potential of a thin film of K$^+$A$^-$ on K$^{2+}$A$^{2-}$
varies with its thickness.

Ten years later Stranski and Krastanov extended the considerations of the same
model by calculating the Gibbs free energies of formation of 2D nuclei of the
first, second, third, etc., monolayers, as well as of two and four 
monolayers-thick
2D nuclei.\cite{Stranski39} It turned out that 2D nuclei of the first monolayer
can be formed at a vapor pressure $P$ which is larger than $P_1$, but smaller
than $P_0$. This means that the first monolayer can be deposited at {\it
undersaturation} with respect to the bulk crystal for the reasons given above. 
The work of formation of 2D nuclei of the second monolayer is very large 
but that of
2D nuclei consisting of two monolayers belonging to the second and third level,
(or in fact three-dimensional), is much smaller. The reason is that the chemical
potential of a bilayer deposited on the first monolayer is lower than that of a
single monolayer but still higher than $P_0$. It was found that the chemical
potential of the bilayer is equal to the arithmetic average of the chemical
potentials of the second monolayer (larger than the bulk chemical potential)
and the third monolayer (slightly smaller than the bulk chemical potential).
This means that the formation of doubly high 2D nuclei requires a {\it
supersaturation}. Note that in the original study of Stranski and Krastanov the
bilayer nuclei which form on the first stable monolayer are two-dimensional.
This means that they should grow laterally and not as 3D islands in length and
height.\cite{Kaischew81}

\begin{figure}[h]
\includegraphics*[width=8.5cm]{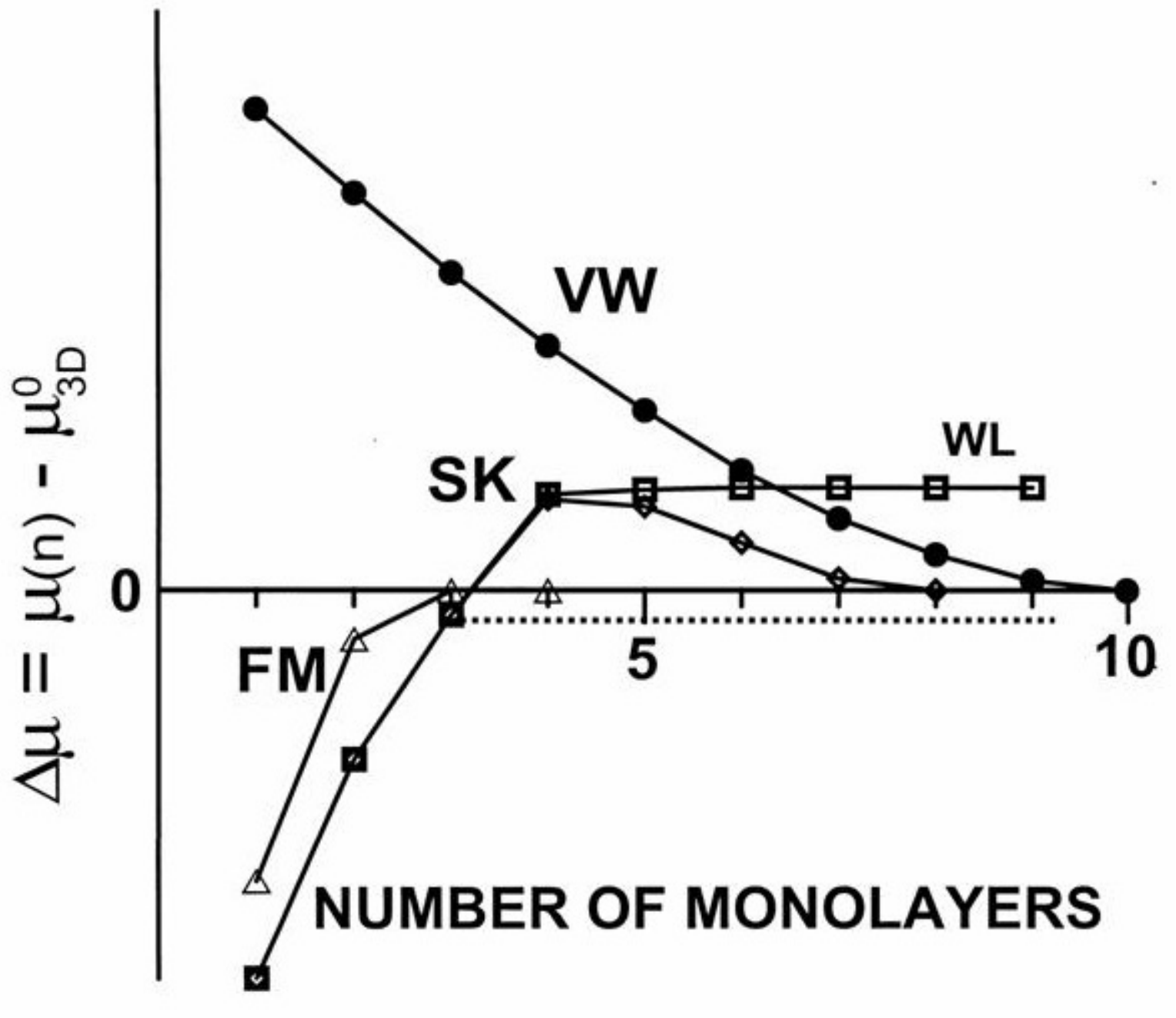}
\caption{\label{tdf} Illustration of the dependence of the chemical potential
of the overlayer on the film thickness in number of monolayers for the three
modes of growth: Volmer-Weber (VW), Frank-van der Merwe (FM) and
Stranski-Krastanov (SK). The upper straight line denoted by WL gives the
chemical potential of the unstable wetting layer (a monolayer in excess which
will be transformed into 3D islands), whereas the lower dotted line gives the
chemical potential of the uppermost monolayer belonging to the stable wetting
layer. (J. E. Prieto, I. Markov, Phys. Rev. B 66, 073408 (2002)). By
permission of the American Physical Society.}
\end{figure}

We now consider a crystal $B$ with lattice parameter $b$ on the surface of a
crystal $A$ with lattice parameter $a$. Contrary to the chemical potential,
$\mu^0_{\rm 3D}$, of the bulk crystal $B$, the chemical potential of the thin
film, $\mu (n)$ will depend on the film thickness measured in number of
monolayers $n$ for two reasons: The first one is that the attraction of the
atoms of the consecutive monolayers by the substrate decreases with
increasing distance from the interface. A second source of dependence on the
film thickness is the mechanism of relaxation of the lattice misfit, either by
introduction of misfit dislocations, by alloying or by the film growing
pseudomorphically with the substrate.

The dependence of the thermodynamic driving force
$\Delta \mu = \mu (n) - \mu ^0_{\rm 3D}$ is shown in Fig.\ \ref{tdf} as a
function of the film thickness measured in number of monolayers.\cite{Prieto02}
These dependences follow from the detailed consideration of the thickness
variation of the film chemical potential\cite{Markov17}
\begin{equation}\label{muensigma}
\mu (n) = \mu^0_{\rm 3D} + a^2(\sigma + \sigma_{\rm i} -\sigma_{\rm s}) +
\varepsilon_{\rm e}(f)
\end{equation}
where $\varepsilon_{\rm e}(f)$ is the homogeneous strain energy per atom stored
in a separate monolayer of crystal $B$ (we assume that the first layers
wetting the substrate are equally strained), and $f = (b-a)/a$ is the lattice
misfit.

Equation (\ref{muensigma}) can be written in the form
\begin{equation}\label{Deltamu}
\Delta \mu = 2\sigma a^2 \Phi + \varepsilon_{\rm e}(f)
\end{equation}
where
\begin{equation}
\Phi = \frac{\sigma + \sigma_{\rm i} -\sigma_{\rm s}}{2\sigma} = 1 -
\frac{\beta}{2\sigma}
\end{equation}
is the so called {\it wetting parameter} or {\it wetting function} with $\beta$
the specific adhesion energy. In fact it is equal to the equilibrium aspect
ratio $h/l$ of a crystal on an unlike substrate.\cite{Bauer58} In terms
of binding energies the wetting parameter reads $\Phi = 1 -
\psi^{\prime}/\psi$.\cite{Kaischew50,Kaischew60} As noted above both expressions
are indentical.\cite{Markov17} Note that the expression for the wetting
parameter given above is derived assuming that the lattice misfit does not
affect the adhesion energy $\beta$ ($=\psi^{\prime}/a^2$).

Let us now consider Fig.\ \ref{tdf} in more detail. In the case of VW growth
($\Phi > 0$) $\Delta \mu$ tends asymptotically to zero from above with
increasing film thickness. The slope $d\Delta \mu /dn$ is negative. In the case
of FM growth ($\Phi \le 0, f \approx 0$), $\Delta \mu$ tends asymptotically to
zero with increasing film thickness from below. The slope $d\Delta \mu /dn$ is
positive. The third important case is
$\Phi < 0$ and $f \neq 0$. As long as the absolute value of $2\sigma a^2\Phi$ in
Eq. (\ref{Deltamu}) is larger than $\varepsilon_{\rm e}(f)$, the film will grow
in a layer-by-layer mode as in the FM case. As growth proceeds, the energetic
influence of the substrate will diminish, and at some thickness the negative
term $2\sigma a^2\Phi$ will become smaller than the positive strain energy
$\varepsilon_{\rm e}(f)$. $\Delta \mu$ will become positive and the next
monolayer will become unstable against 3D islanding.

Let us now consider the problem in terms of equilibrium vapor pressures although
the connection between the latter and the chemical potentials is straightforward
($\mu \propto {\rm ln}P$). When $\mu (1) < \mu_{\rm 3D}^0$ the first monolayer
will be deposited at a vapor pressure higher than the equilibrium vapor
pressure, $P_1$, of the first monolayer but smaller than the equilibrium vapor
pressure, $P_0$, of the bulk crystal. The same is valid for all monolayers
belonging to the stable wetting layer as long as $\mu (n) < \mu_{\rm 3D}^0$. It
follows that stable monolayers including the uppermost one will be deposited at
vapor pressures $P_1 < P < P_0$ or, in other words, at {\it undersaturation}
with respect to the bulk crystal. The deposition of 3D islands will take place
at $\mu (n) > \mu_{\rm 3D}^0$, or at a {\it supersaturation} with respect to the
bulk crystal.

Hence, in the case of the SK mode of growth, the equilibrium vapor 
pressure of the
uppermost monolayer which belongs to the stable wetting layer (the dotted line
in Fig.\ \ref{tdf}) is lower than the equilibrium vapor pressure, $P_0$ of the
bulk crystal, whereas the 3D islands on top are in equilibrium with a vapor
pressure (the upper straight line denoted by WL in Fig.\ \ref{tdf}) which is
higher than $P_0$. The dividing line is $\Delta \mu = 0$ or $P = P_0$. At this
pressure, the stable wetting layer cannot grow thicker and 3D islands cannot be
formed. Material cannot be transferred from the stable wetting layer to the 3D 
islands as this implies an increase of the free energy of the system. Therefore
it is thermodynamically unfavored. A planar film thicker than the stable 
wetting layer is unstable and the material in excess must aggregate into
3D islands upon annealing. We conclude that the wetting layer and the 3D 
islands are different phases in the sense of Gibbs (``homogeneous parts in a
heterogeneous system")\cite{Gibbs28} which can never be in equilibrium with 
each other.

\section{Effect of lattice misfit on the film-substrate adhesion}

It is instructive to consider first in some detail the one-dimensional
model of Frank and van der Merwe of a finite chain of atoms in the sinusoidal
potential field exerted by a rigid substrate.\cite{Frank49,Frank491} We will
employ anharmonic bonds (steeper repulsion and weaker attraction branches of the
interatomic potential) connecting the atoms instead of the harmonic
approximation used by the authors.\cite{Markov84,Markov841} Such a consideration
will give us valuable information concerning the coherent SK growth mode.

\begin{figure}[htb]
\includegraphics*[width=8.5cm]{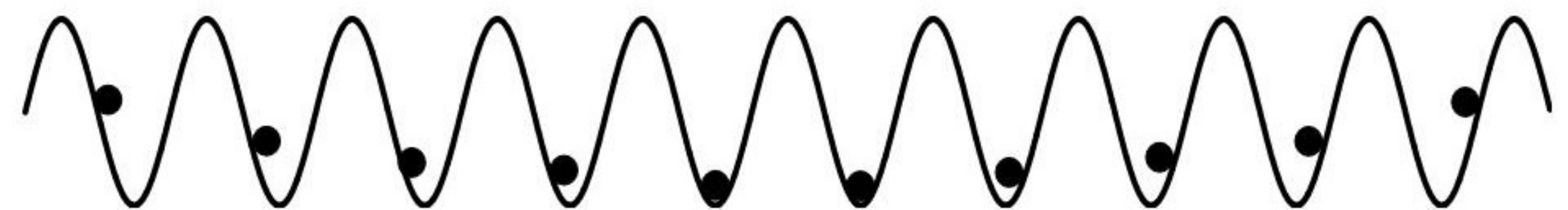}
\caption{\label{FMchain} Schematic representation of a finite crystal
represented by the 1D chain model of Frank and van der
Merwe.\cite{Frank49,Frank491}}
\end{figure}

In the previous section we considered the wetting parameter and the lattice
misfit as independent variables. Frank and van der Merwe found that the end
atoms in the chain are displaced from their sites in the bottoms of the
potential troughs of the substrate. These displacements lead to two effects.
First, the atoms close to the chain ends adhere more weakly to the substrate as
compared with the atoms at the center, the chain as a whole loses contact with
the wetting layer underneath and an effective positive wetting parameter results
(Fig.\ \ref{FMchain}). This appears as the thermodynamic driving force for
the 2D-3D transformation. If all atoms are in the bottoms of the potential
troughs, the wetting parameter will be precisely equal to zero and 3D islanding
will be impossible. Note that if the chain is infinitely long the atoms will not
be displaced and the wetting parameter will be again equal to zero. This is
in fact the case for the monolayers belonging to the wetting layer.

\begin{figure}[htb]
\includegraphics*[width=7.5cm]{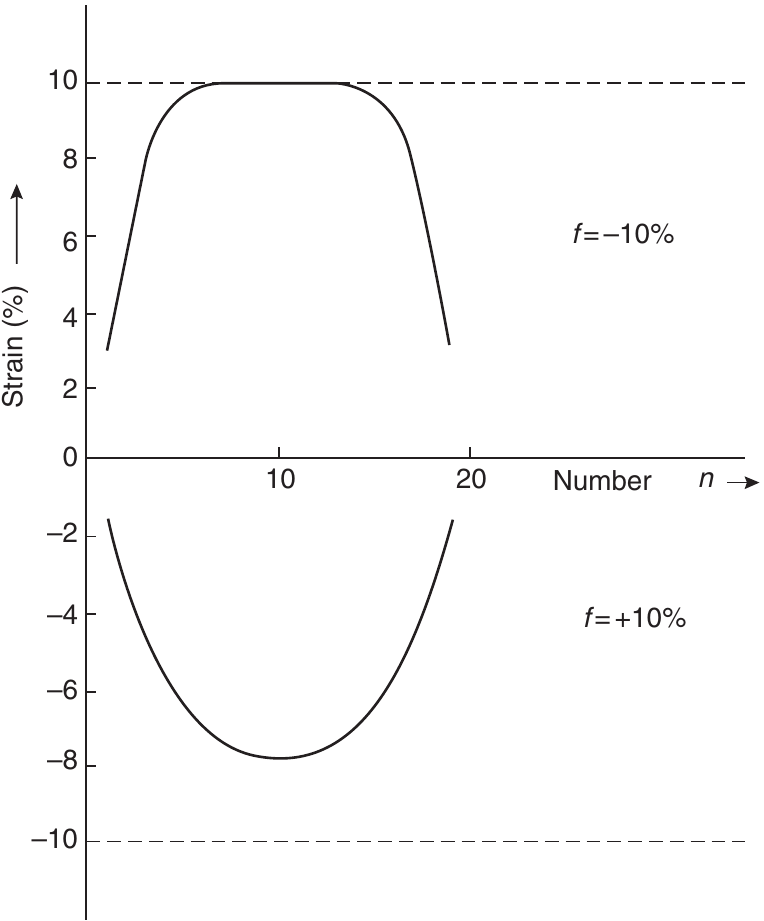}
\caption{\label{strains} Distribution of the strain in monolayer height
compressed ($f = 0.10$) and tensile ($f = - 0.10$) chains.}
\end{figure}

Thus the finite 2D islands do not wet completely the substrate and at some
critical size, they become unstable against bilayer or multilayer (3D) islands
as discussed below. Note that beyond the wetting layer
the energetic influence of the unlike substrate disappears and this makes the
2D-3D transformation possible. Such a transformation in the layers belonging
to the wetting layer is impossible for thermodynamic reasons: the attraction of
the atoms from the unlike substrate prevails over the strain energy per bond.
Note also that the atoms in the tensile chain adhere more strongly to the
wetting layer since more bonds are strained to fit it. On the contrary, bonds
located further away from the ends in the case of compressed chains are also
partially relaxed. Second, the bonds very close to the chain ends are 
maximally relaxed. If the chain consists of $N+1$ atoms connected by $N$ 
bonds, the hypothetical 0-th and ($N$+1)-st bonds would be completely 
unstrained (Fig.\ \ref{strains}).\cite{Frank491,Markov84,Korutcheva00}

We support our thermodynamic considerations by numerical calculations making
use of a simple minimization procedure. We make use of two models. The first
is the same atomistic model in 1+1 dimensions (length + height) as in Refs.
(\onlinecite{Korutcheva00}) and (\onlinecite{Ratsch93}) for fast qualitative
calculations. The 3D islands are represented by linear chains of atoms stacked
one upon the other. The island height is considered as a discrete variable
which increases by unity starting from one. The second model is the more
realistic (2+1)-dimensional construction [(length + width) + height]. The
substrate (the wetting layer) in both cases is assumed to be rigid. In all cases
we consider a crystalline film with an fcc lattice and (100) orientation 
at zero temperature. The atoms interact through an anharmonic pair-wise
potential\cite{Markov93}
\begin{equation}\label{potential}
V(x) = V_0 \Bigl[\frac{\nu}{\mu - \nu}e^{-\mu (r-b)} - \frac{\mu}{\mu -
\nu}e^{-\nu (r-b)}\Bigr]
\end{equation}
which, in spite of its simplicity, includes all necessary features to describe
real materials (its strength and anharmonicity are governed by the constants
$\mu$ and $\nu$, $\mu > \nu$). In the case of $\mu = 2\nu$ it turns into the
familiar Morse potential. We consider interactions only in the first
coordination sphere in order to mimic the directional bonds that are
characteristic of elemental semiconductors.\cite{Tersoff86}

Let us emphasize once again that our system consists of a substrate crystal $A$,
a stable wetting layer of crystal $B$, and 2D or 3D islands of crystal $B$ on
top of the wetting layer. When studying the formation of 3D islands on the
wetting layer we consider the latter as a substrate. Thus we consider the growth
of $B$ on strained $B$. As we study the initial stages of the formation of 3D
islands, they are assumed to be sufficiently small and coherent with respect 
to the wetting layer. 
In other words, we study the coherent (dislocationless) SK growth. In
addition, we assume that the wetting layer is pseudomorphous with the substrate
crystal $A$, i.e. the separate monolayers forming the wetting layer are equally
strained and possess the same interatomic spacing as $A$.

\begin{figure}[htb]
\includegraphics*[width=7.5cm]{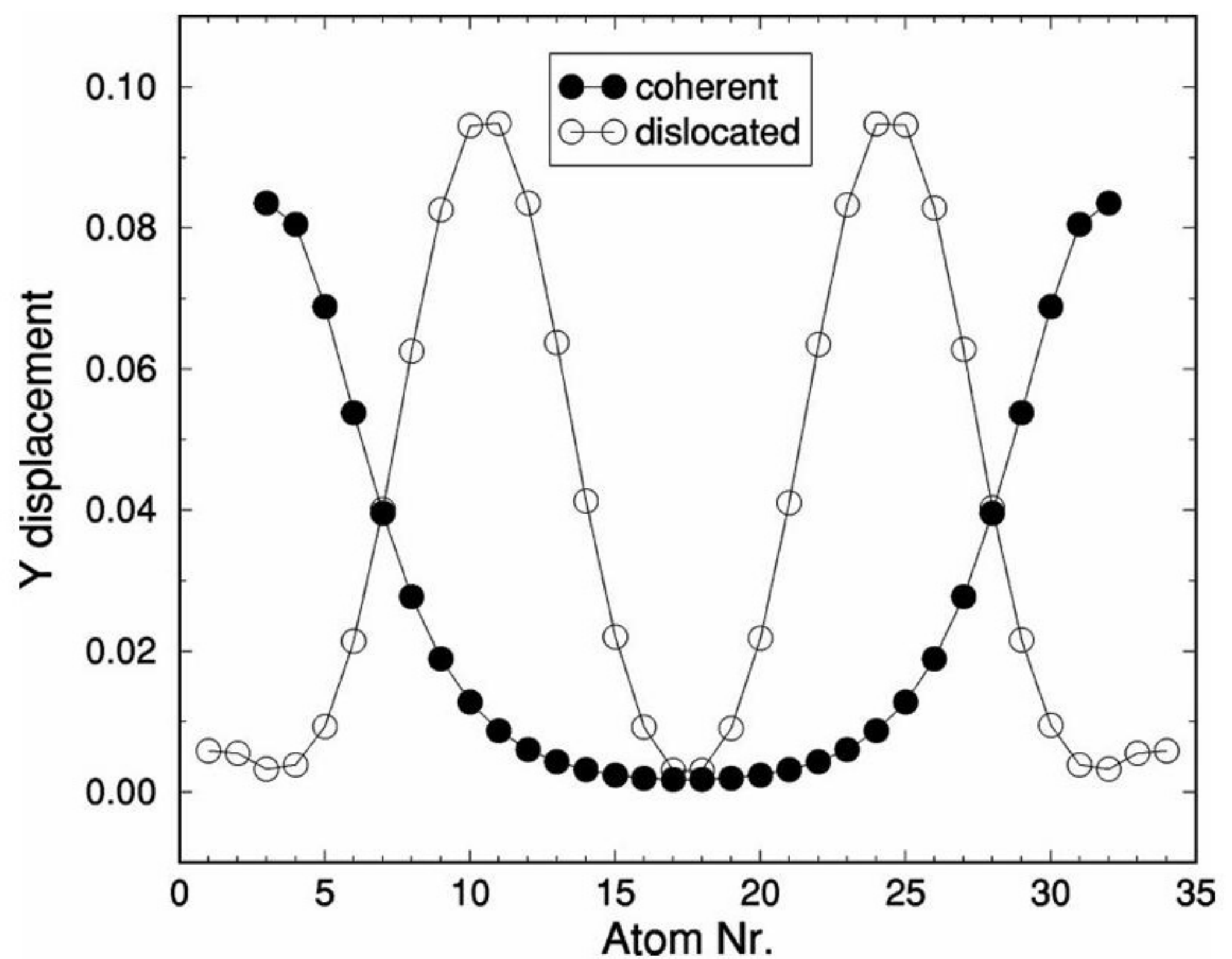}
\caption{\label{verticals} Vertical displacements of the atoms of the base chain
of a coherent ($\bullet$) and a dislocated ($\circ$), 3-monolayer-thick island.
This is an illustration of the reduced adhesion of the islands to the wetting
layer. The displacements are given in units of the lattice parameter of the
wetting layer (which is equal to that of the substrate crystal) and are measured
from the bottoms of the potential troughs of the wetting layer. The misfit $f$
amounts to 7\% and the islands contain 30 and 34 atoms in their base chains,
respectively. (J. E. Prieto, I. Markov, Phys. Rev. B 66, 073408 (2002)).
By permission of the American Physical Society.}
\end{figure}

\begin{figure}[htb]
\includegraphics*[width=7.5cm]{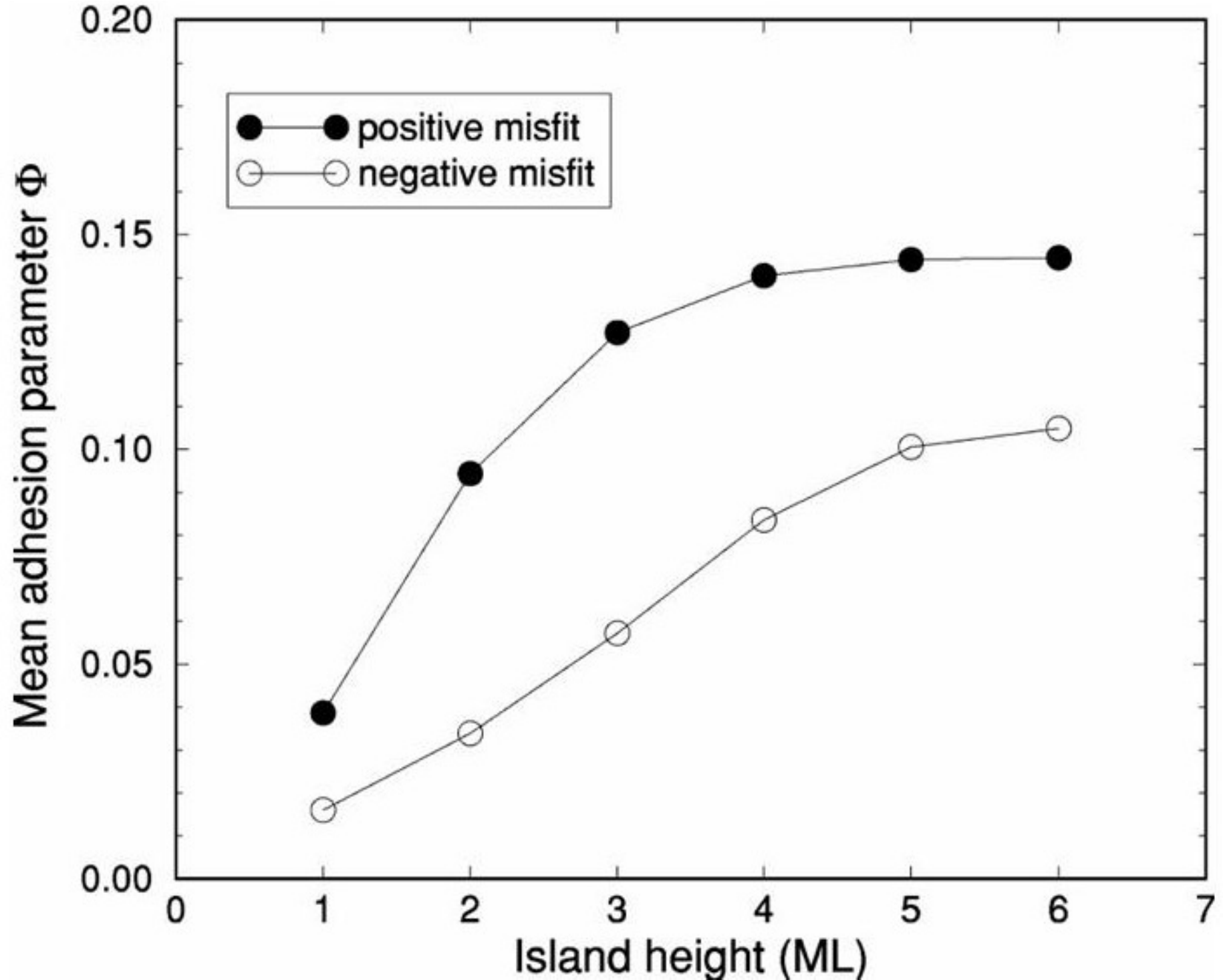}
\caption{\label{phi-thickness} Mean adhesion parameter $\Phi$ as a function of
the island height in number of monolayers for positive and negative values of
the misfit of absolute value of 7\%. The base chain consists of 14 atoms. (J. E.
Prieto, I. Markov, Phys. Rev. B {\bf 66}, 073408 (2002)). By permission of the
American Physical Society.}
\end{figure}

Fig.\ \ref{verticals} illustrates the differences (and similarities) between the
coherent and incoherent (dislocated) SK growth modes. In the first case the
end atoms are displaced vertically in the potential troughs of the substrate.
In the second case the atoms in the cores of the dislocations are displaced
in a similar way. As seen, in both cases the 3D islands (and the 2D islands as
well, see above) lose contact with the substrate. Note that the dislocated
island is a little bit longer (34 atoms) than the coherent one (30 atoms). This
means that there is a critical size for the transformation of coherent into
dislocated islands. The mean adhesion parameter $\Phi$ increases with the
island's height and saturates beyond several monolayers (Fig.\
\ref{phi-thickness}). Here $\Phi$ is calculated as the average adhesion
energy of the atoms between the base chain and the wetting layer at the given
value of the misfit minus the corresponding value for zero misfit. This is a
very important result as it shows that thicker islands behave effectively as
``stiffer" ones, i.e. with stronger interatomic bonds, as discussed below. This
is directly connected with the multilayer mechanism of transformation of
monolayer to multilayer islands.

\begin{figure}[htb]
\includegraphics*[width=7.5cm]{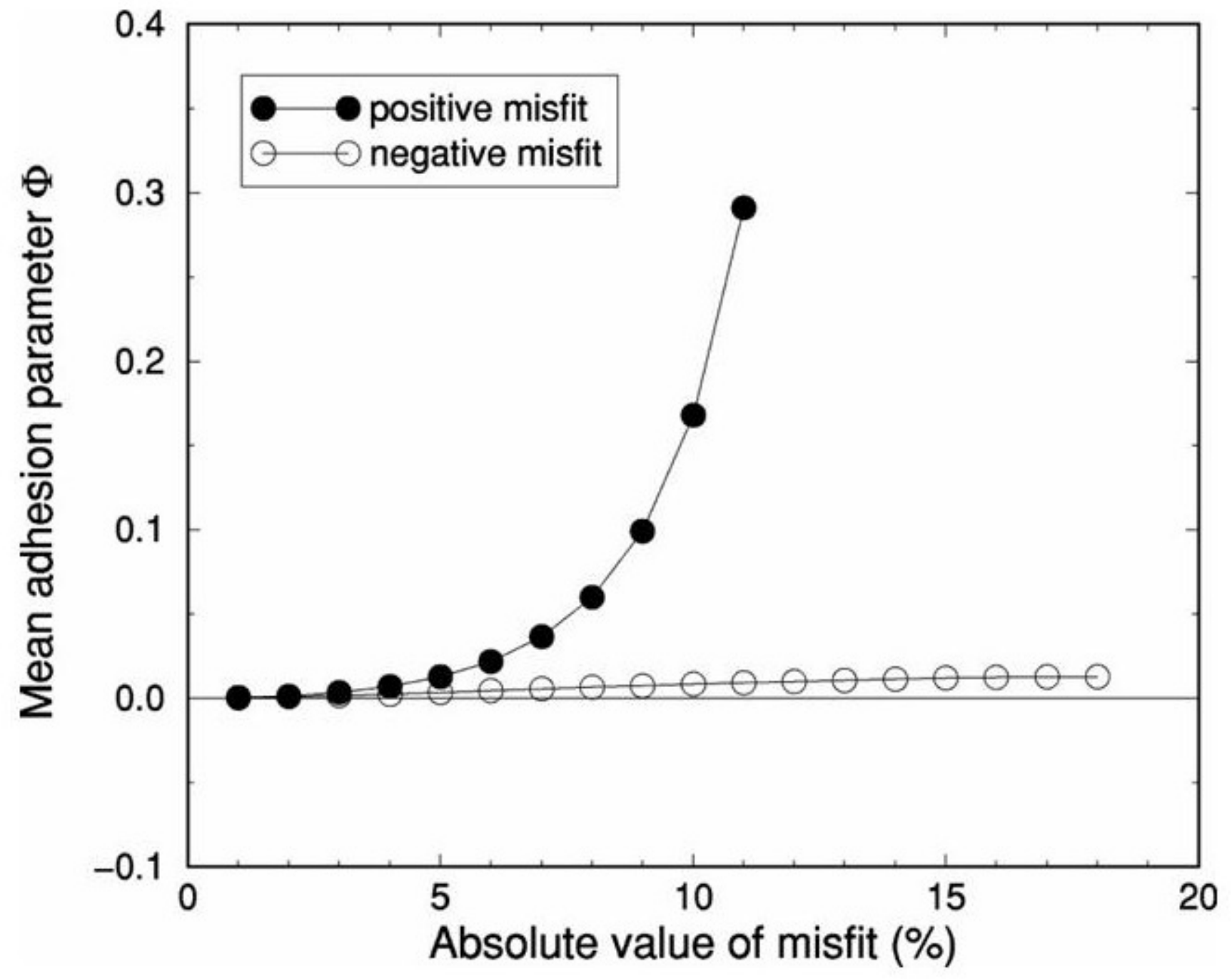}
\caption{\label{Phi vs misfit} Mean adhesion parameter of one monolayer thick
coherent islands as a function of the lattice misfit. The islands consist of 20
atoms. Data for both compressed ($\bullet$) and tensile ($\circ$) islands
are shown in one and the same quadrant for easier comparison. (J. E. Prieto, I.
Markov, Phys. Rev. B {\bf 66}, 073408 (2002)). By permission of the American
Physical Society.}
\end{figure}

Fig.\ \ref{Phi vs misfit} shows the dependence of the mean adhesion parameter
on the lattice misfit in the case of coherent 3D islands. In fact, as discussed
above, this is the dependence of the thermodynamic driving force for 3D island
formation. As seen it is much larger in compressed islands than in 
tensile ones. The wetting parameter of tensile islands remains very small
whereas that of compressed islands increases steeply for values of the lattice
misfit beyond approximately 5\%. The same qualitative results have been 
obtained in the case of a (2+1) dimensional model.\cite{Prieto05}

We conclude that the lattice misfit strongly affects the adhesion of the 3D
islands on the strained wetting layer. The mean adhesion parameter and in turn
the thermodynamic driving force for 3D island formation beyond the wetting
layer are much larger for compressed islands than for tensile ones, for which
$\Phi$ remains close to zero. This means that tensile films display a much 
smaller tendency (if any) to Stranski-Krastanov growth as compared with 
compressed ones.

\section{Thickness of the stable wetting layer}

We can now write Eq. (\ref{muensigma}) in terms of the binding energies $\psi$
(cohesion) and $\psi^{\prime}$ (adhesion) in the form
\begin{equation}\label{muenpsi}
\mu (n) = \mu^0_{\rm 3D} + \psi - \psi^{\prime}(n) + \varepsilon_{\rm e}(f)
\end{equation}
where
\begin{equation}\label{strain}
 \varepsilon_{\rm e}(f) = 2G_{\rm b}a^2 h f^2 \frac{1+\nu_b}{1-\nu_b}
\end{equation}
is the energy of homogeneous strain per atom stored in each separate monolayer,
$h$ is the thickness of a monolayer and $G_{\rm b}$ and $\nu_{\rm b}$ are the
shear modulus and the Poisson ratio of the crystal $B$, respectively.
The contribution of the misfit dislocations is omitted since it is
expected that they will be introduced at a thickness larger than that
of the stable wetting layer.

We have to find the dependence of the adhesion energy $\psi^{\prime}(n)$ on 
the thickness $n$. This is a result of the decrease of the adhesion with
the distance from the substrate surface. The first monolayer is most strongly
attracted by the substrate, the second more weakly, the third very weakly, and
the fourth monolayer most probably will not feel the presence of the substrate.
Two types of $\psi^{\prime}(n)$ dependences are generally accepted;
an inverse cubic $n^{-3}$ dependence for van der Waals bonding (noble gases)
and an exponential decay $e^{-n/n_0}$ for metallic and covalent bonding in
semiconductors.\cite{Muller96} It is worth noting that an empirical interatomic
potential was devised for the properties of Si on the basis of a Morse-like
pair-wise potential.\cite{Tersoff86}

The inverse cubic dependence $n^{-3}$ follows from a pair-wise interaction of
the Lennard-Jones 6-12 type between the adsorbate and the
substrate\cite{Dash77,Mutaftschiev01}
\begin{equation}\label{psien}
\psi^{\prime}(n) = \psi - \frac{\psi - \psi^{\prime}_1}{n^3}
\end{equation}
where $\psi^{\prime}_1$ is the energy of desorption of an atom belonging to
the first monolayer of the wetting layer from the unlike substrate at a
coverage tending to zero.

Combining Eqs. (\ref{muenpsi}), (\ref{strain}) and (\ref{psien}) under the
condition $\mu (n) = \mu^0_{\rm 3D}$ gives for the thickness of the stable
wetting layer
\begin{equation}
n = \Bigl(\frac{\psi}{\varepsilon_{\rm e}}|\Phi|\Bigr)^{1/3}
\end{equation}
where the absolute value, $|\Phi|$, of the wetting parameter must be taken.

In the case of an exponential decay of the influence of the substrate, the
thickness of the stable wetting layer reads\cite{Muller96}
\begin{equation}
\frac{n}{n_0} = {\rm ln}\Bigl(\frac{\psi}{\varepsilon_{\rm e}}|\Phi|\Bigr)
\end{equation}
where $n_0$ is a parameter of order unity that can be determined 
by comparison with experiments.\cite{Muller96}

It is of interest to compare both expressions for $n$. In the case of
deposition of Ge on Si(001) $\varepsilon_{\rm e} = 0.035$ eV/atom with $G_{\rm
Ge} = 5.64 \times 10^{11}$ dyne/cm$^2$, $\nu = 0.2$, $a = 3.84\AA$, $h =
1.4\AA$ and $f = 0.041$. Assuming that the Ge atoms belonging to the
first monolayer
are attracted by the Si substrate with the same force as Si atoms (Ge and
Si have similar chemical properties) $\Phi = - 0.216$, and with $\psi = 1.94$
eV (one half of the enthalpy of evaporation), $\psi |\Phi| /\varepsilon _{\rm
e} = 12.0$. We then obtain $12^{1/3} = 2.3$, and ${\rm ln}12 = 2.48$. In other
words we obtain two close values, similarly low when compared with the
experimentally found value of 3 monolayers. This coincidence looks 
somewhat strange bearing in mind the different physics behind both 
expressions. In addition the parameter $n_0$ remains unknown from the 
theoretical point of view.

The formulas above seem to imply that the thickness of the stable wetting layer
tends to infinity when the lattice misfit and in turn the strain energy tend to
zero. It is then noteworthy to mention that two reasons oppose this conclusion.
First, as shown below, the 3D islanding is only possible at values of the misfit
larger than some critical value. Even if this would not be the case, the
decrease of the lattice misfit should lead to a transition from the
Stranski-Krastanov mode to the Frank-van der Merwe mode of growth.

Let us consider a wetting layer consisting, say, of three equally strained
monolayers, pseudomorphic with the substrate. The closer to the substrate, 
the more strongly a given monolayer is attracted by it. The
decrease of the binding of each layer to the substrate is a discontinuous
(step-like) function. It follows that the equilibrium vapor pressure of the
consecutive monolayers (or the chemical potential) is an increasing step-like
function of the number of monolayers but all these values are lower than the
equilibrium vapor pressure of the bulk deposit crystal.\cite{Grabow88} The
growth of the next monolayer begins after the completion of the previous one as
the chemical potential of the latter is smaller. Hence, each monolayer which
belongs to the stable wetting layer is a distinct two-dimensional phase, and the
formation of each new monolayer is a phase transition of first
order.\cite{Muller96} Thus, in the particular case of Ge/Si(001) we have 4
Ge-containing phases, three separate monolayers belonging to the stable wetting
layer plus the 3D islands, all four phases possessing different equilibrium 
vapor pressures.

We conclude that the thickness of the stable wetting layer increases as
expected with increasing absolute value of the wetting parameter and
with decreasing lattice misfit, but must remain always in the range
of action of the interatomic forces, i.e., not more than 3-4
monolayers. Note that, as discussed above, the increase of the lattice misfit
leads to an increase of the wetting parameter, so both parameters have
opposite effects on the thickness of the wetting layer.

It follows from the above considerations that the number of the monolayers 
belonging to the
stable wetting layer must be an integer. The thickness of the wetting layer
has been determined in numerous papers as the thickness for the onset of 3D
islanding. In all cases non-integer values have been established. Thus the
values of 3.7 MLs in the case of Ge on Si(001),\cite{Sunamura95} 3.1-3.4 MLs in
the case of Ge/Si(111),\cite{Grimm16} 1.4 MLs\cite{Saint-Girous02} and 1.75
MLs\cite{Moison94} in the InAs/GaAs system, were measured from the onset of
3D islands formation. Borgi {\em et al.} found that the onset of 3D
islanding depends both on the temperature and the substrate orientation
in the case of deposition of InP on the (100) and (111)$_{\rm A,B}$
surfaces of GaP.\cite{Borgi99}

However, in all cases the amount of the material deposited in excess of the
corresponding integer number of monolayers tends steeply to zero after 
the onset of formation of 3D islands.\cite{Sunamura95,Moison94} This 
unambiguously
shows that the amount of deposit in excess is consumed by the 3D islands.
Thus it is the integer number of monolayers that constitute the stable wetting
layer. If we divide the excess material by the number of the 3D islands we
can estimate the critical size of the monolayer islands which appear as
precursors of the 3D islands, as shown below.

\section{Stability of mono- and multilayer islands}

\begin{figure}[htb]
\includegraphics*[width=7.5cm]{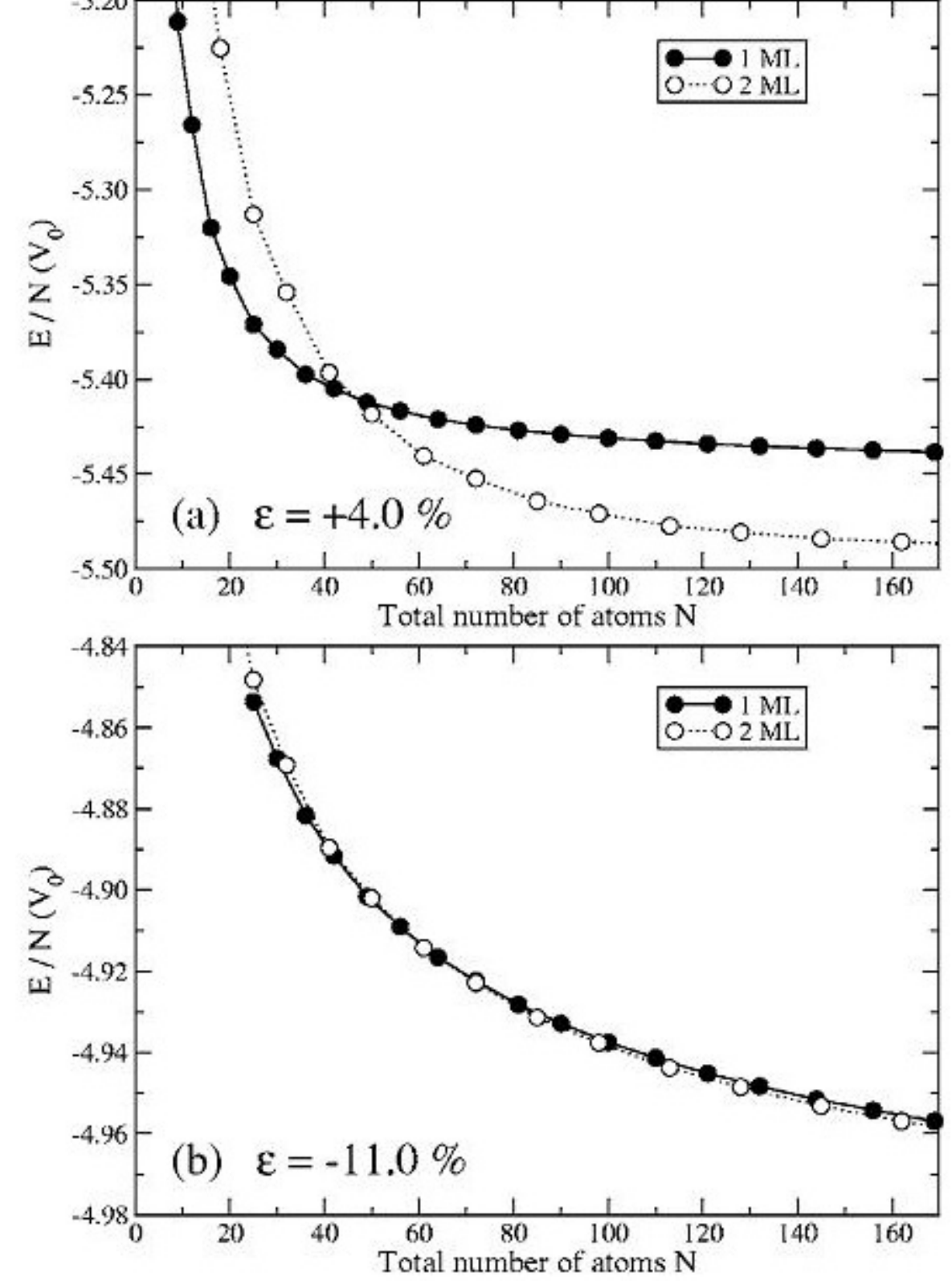}
\caption{\label{mono-bilayer} Total energy per atom of mono- and bilayer
islands at (a) positive and (b) negative values of the misfit as a function of
the total number of atoms. The atoms interact through the potential
(\ref{potential}) with $\mu = 16$ and $\nu = 14$. (J. E. Prieto, I. Markov,
Phys. Rev. B 72, 205412 (2005)). By permission of the American Physical
Society.}
\end{figure}

Following the approach in Ref. (\onlinecite{Stoyanov82}) we plot the binding
energies per atom of monolayer, bilayer, trilayer islands, etc., as a function
of the total number of atoms making use of the more realistic (2+1)-D
model.\cite{Prieto05} Fig.\ \ref{mono-bilayer} shows that the total energies of
monolayer and bilayer islands under tensile stress are very close to each other
irrespective of the larger absolute value of the misfit (-11\%) compared with
the corresponding behavior of compressed islands. Nevertheless, in both cases we
observe a critical size, $N_{\rm 12}$, beyond which bilayer islands become
energetically favored. Three-layer islands (not shown) become energetically
favored beyond a critical size $N_{\rm 23} > N_{\rm 12}$, etc. Note that the
crossover at positive misfit is much more pronounced in spite of the
smaller absolute value of the misfit.

Thus we can expect that initially 2D islands are formed on the stable wetting
layer, which beyond a critical size, $N_{\rm 12}$, become unstable against
bilayer islands. The latter become unstable against trilayer islands beyond a
critical size, $N_{\rm 23}$, etc. Thus the monolayer islands appear as
necessary precursors of the 3D islands.\cite{Priester95,Chen96} Voigtl\"ander
and Zinner observed by STM that faceted 3D Ge
islands are formed at the same locations on a Si(111) surface at which 2D
islands were observed in the initial stage of deposition immediately after
exceeding the critical thickness of the wetting layer.\cite{Voigt93} Ebiko
{\it et al.} found that the scaling function of the volume distribution of 3D
InAs quantum dots on the surface of GaAs coincides with the scaling function for
2D submonolayer homoepitaxy with critical nucleus size $i^* = 1$.\cite{Ebiko99}
Note that in submonolayer homoepitaxy the size of the 2D nucleus is defined as
the compact stable cluster minus an atom and depends strongly on the
temperature. Thus in the case of an fcc (100) surface the nucleus can 
consist of 3 atoms and the stable cluster of 4 atoms forming a
square.\cite{Walton69} As will be shown below, the critical nucleus which gives
rise to a new monolayer in the 2D-3D transformation, and the driving force for
its formation is the lattice misfit, is defined in a different way.
XXXX

\begin{figure}[htb]
\includegraphics*[width=7.5cm]{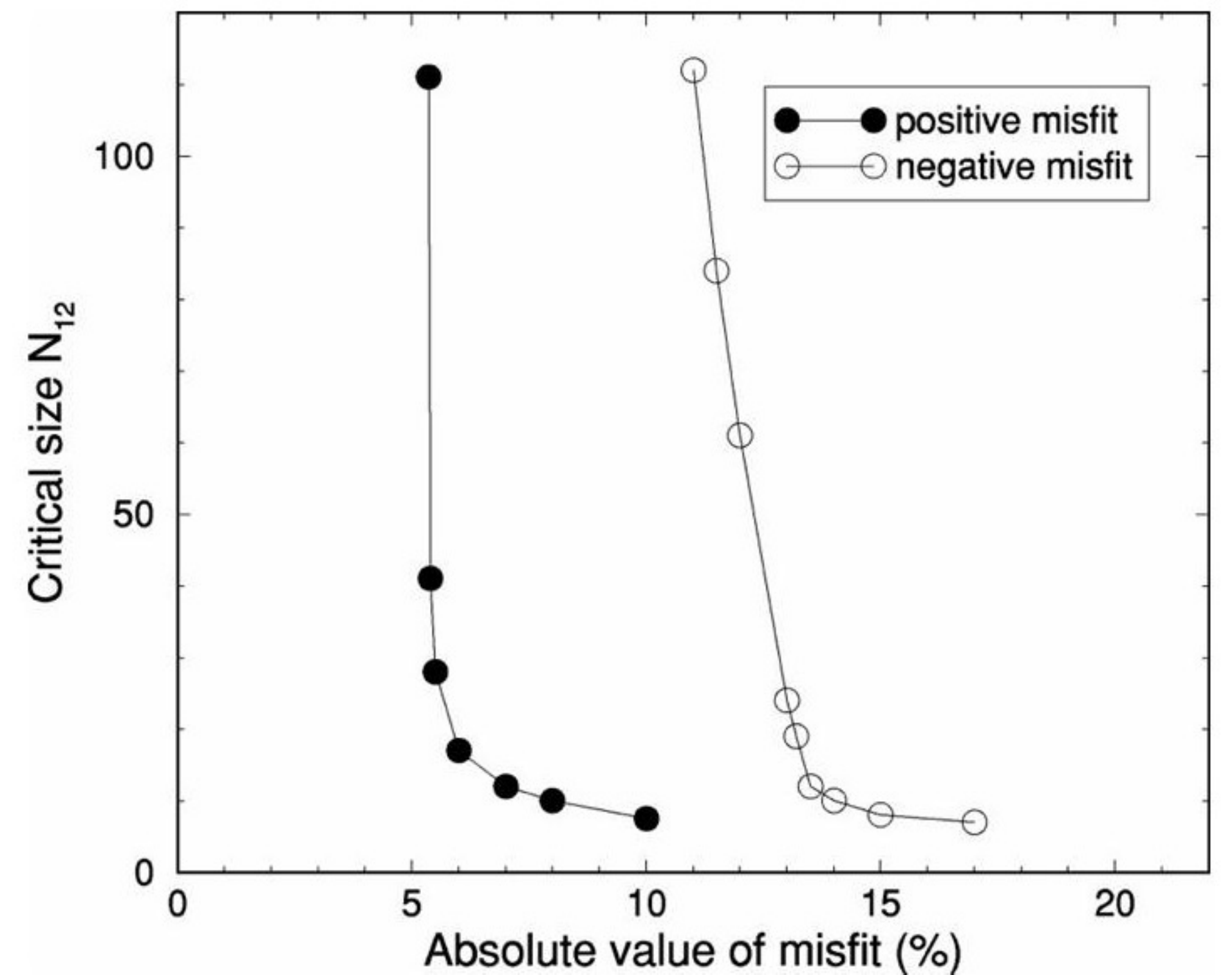}
\caption{\label{en12} The critical size, $N_{\rm 12}$, (in number of atoms)
as a function of the lattice misfit for positive and negative values of the
latter. The curves are calculated by making use of the (1+1)D model. The curves
are shown in one quadrant for easier comparison. (J. E. Prieto, I. Markov, Phys.
Rev. B 66, 073408 (2002)). By permission of the American Physical Society.}
\end{figure}

It turns out that islands of a given thickness $t$ are stable in an interval of
sizes (in number of atoms) between $N_{\rm t-1,t}$ and $N_{\rm t, t+1}$. We then
plot for simplicity the critical size $N_{\rm 12}$ for mono-bilayer
instability, assuming that the bilayer islands can be considered as
three-dimensional. Fig.\ \ref{en12} shows the dependence of the critical size
$N_{\rm 12}$ on the lattice misfit for the cases of compressive and 
tensile overlayers. 
The calculations are performed with the (1+1)D model.\cite{Prieto02} A sharp
increase with decreasing absolute value of the misfit is observed only in the
case of compressed overlayers, where $N_{\rm 12}$ clearly goes to infinity at
some critical misfit $f_{\rm cr}$. It is woth to note that in tensile overlayers
the increase is less sharp and the curve is displaced to larger absolute values
of the misfit. We conclude that coherent 3D islands can be formed at misfits
larger than $f_{\rm cr}$.

\begin{figure}[htb]
\includegraphics*[width=7.5cm]{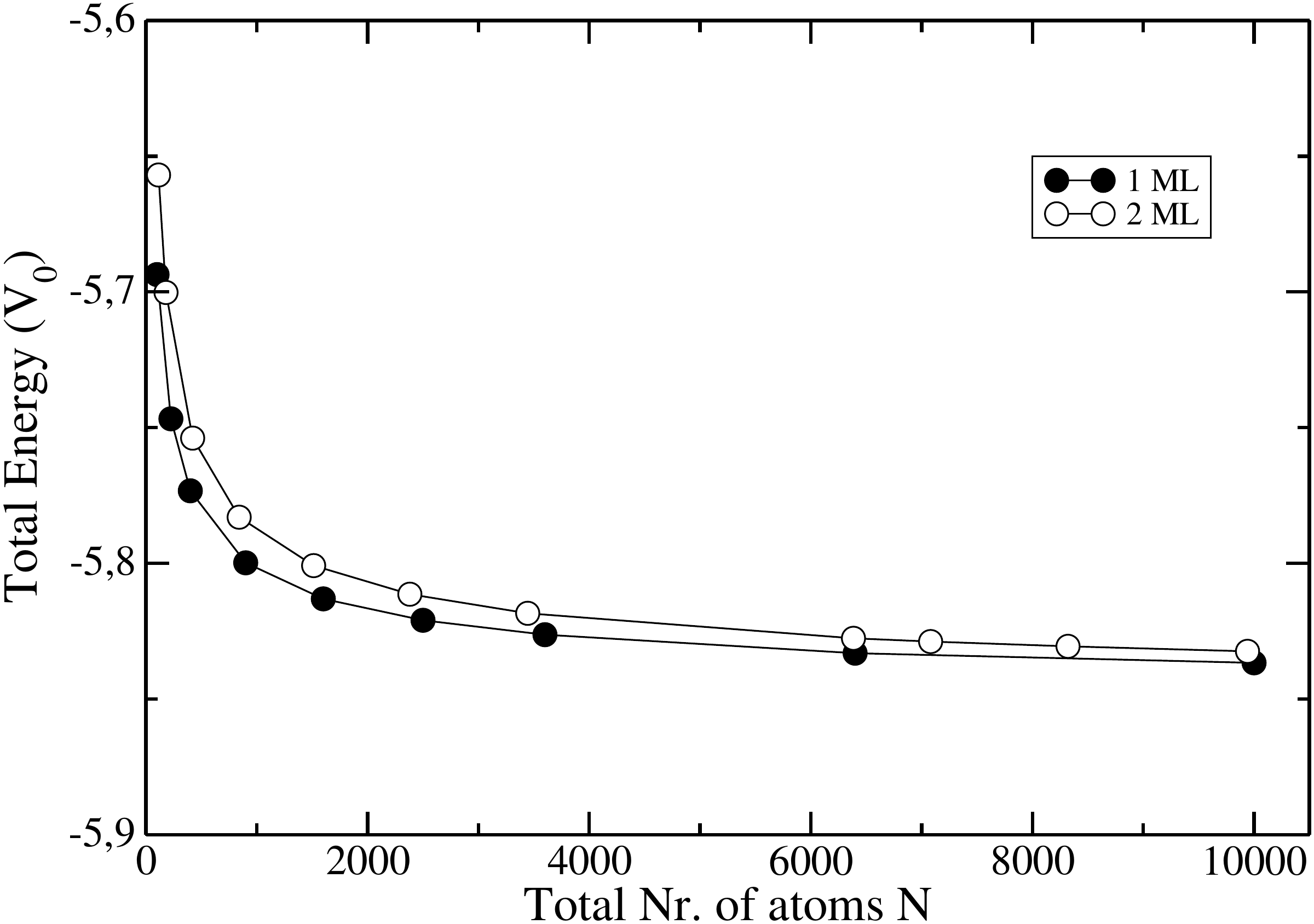}
\caption{\label{smallmisfit} Dependence of the energies of mono- and bilayer
islands in (2+1)D for a misfit of 4\%, smaller than the critical misfit $f_{\rm
cr} = 5.2\%$, ($\mu = 2\nu = 12$).}
\end{figure}

For misfits smaller than $f_{\rm cr}$ the film is expected to continue growing
in a layer-by-layer mode until misfit dislocations are introduced to relax the
strain. This is clearly demonstrated in Fig.\ \ref{smallmisfit} which shows that
monolayer islands in the (2+1)D model remain stable against bilayer ones up to a
number of atoms as large as 10.000. The existence of a critical misfit for
formation of coherent quantum dots does not allow the formation of a stable
wetting layer thicker than the range of action of the interatomic bonding
as discussed above.

\begin{figure}[htb]
\includegraphics*[width=7.5cm]{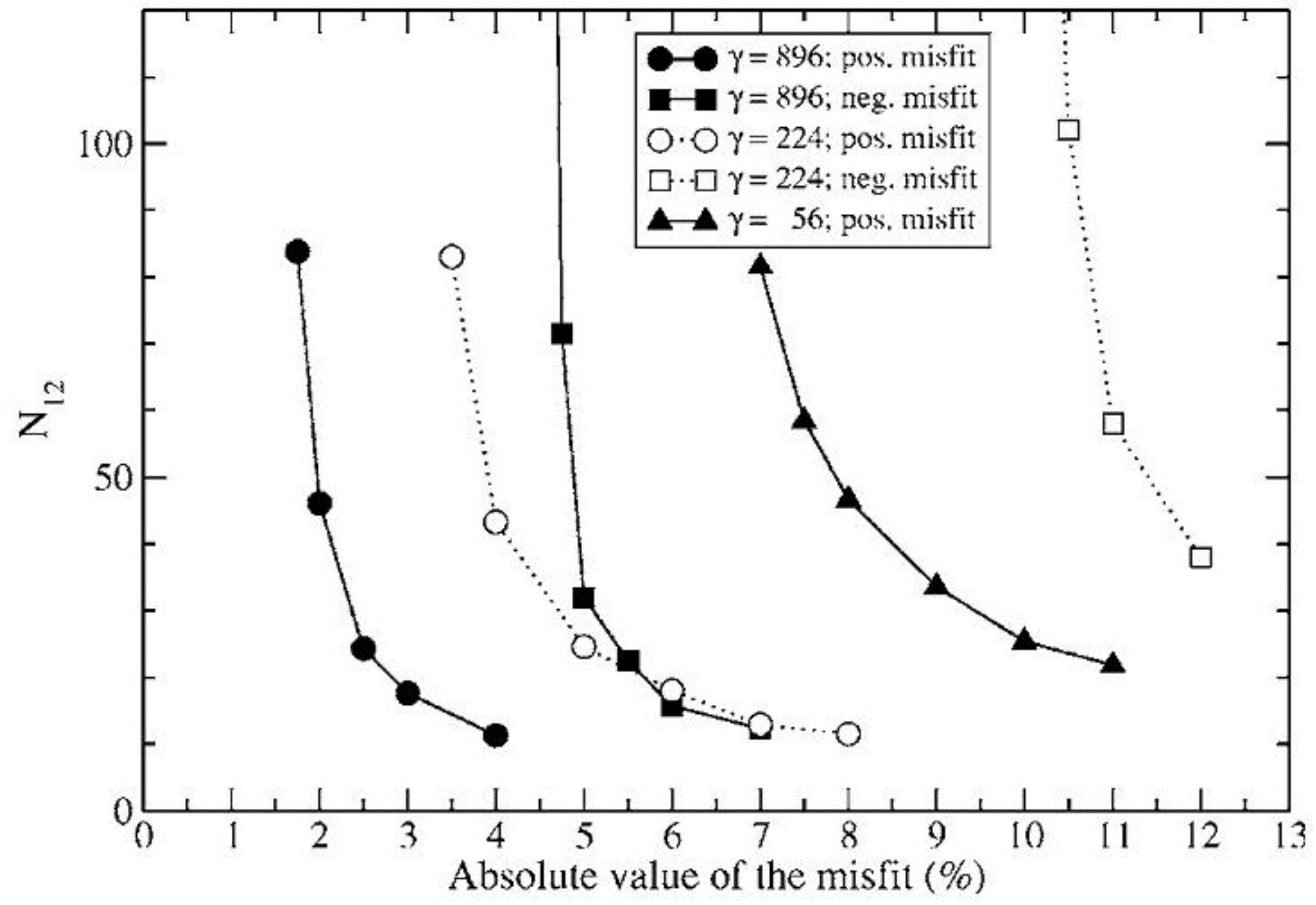}
\caption{\label{en123D} Critical island size $N_{\rm 12}$ as a function of the
lattice misfit at different values of the force constant $\gamma = \mu \nu
V_0$. The (2+1)D model and the potential (\ref{potential}) were used 
with $\mu /\nu = 8/7$. Coherent 3D islanding is favored in tensile epilayers 
only for ``stiff" materials. (J. E. Prieto, I. Markov, Phys. Rev. B 72, 
205412 (2005)). By permission of the American Physical Society.}
\end{figure}

Fig.\ \ref{en123D} shows the misfit dependence of the critical island size
$N_{\rm 12}$ in the (2+1)D model.\cite{Prieto05} Here, a new parameter, the bond
strength or the force constant of the interatomic bonds $\gamma = \mu \nu V_0$,
is varied. Decreasing $\mu$ and $\nu$ in such a way that the ratio $\mu /\nu$ is
kept constant shifts $N_{\rm 12}$ to larger absolute values of the misfit. This
implies in practice the disappearance of $N_{\rm 12}$, since it shifts to
unrealistically high values of the misfit at small values of $\gamma$ in
particular in tensile overlayers whereas $N_{\rm 12}$ exists practically
for all values of $\gamma$ in compressed overlayers.

Another very important result is that in the case of an intermediate value of
$\gamma$ ($\mu = 2\nu = 12$) the monolayer islands are always stable against
bilayer islands ($N_{\rm 12}$ disappears) but $N_{\rm 13}, N_{\rm 14}\hdots$
still exist, where $N_{\rm 13}$ and $N_{\rm 14}$ are the cross points of
the energies of monolayer and three- and four-layer thick islands. At even
smaller values of $\gamma$ the critical values
of $N_{\rm 13}$, $N_{\rm 14}\hdots$ consecutively disappear which leads to the
idea of a novel mechanism of growth of the 3D islands which differs from the
layer-by-layer growth. This mechanism should consist of a direct
transformation of monolayer into multilayer islands by nucleation and lateral
growth of two-dimensional multilayer islands on top of the initial monolayer
islands. This transformation should take place at sizes greater than $N_{\rm
1X}$. Obviously this size should be much greater than $N_{\rm 12}$. We thus
conclude that the 3D islands should form and grow by two distinctive mechanisms,
one for ``stiffer" and the other for ``softer" materials. These two cases 
will be considered separately in more detail below.

\section{Layer-by-layer growth of 3D islands}

The layer-by-layer mechanism of formation and growth of 3D islands was first
suggested by Stoyanov and Markov\cite{Stoyanov82} (see also Ref.
\onlinecite{Markov87}) in the case of the VW growth of an elastically
unstrained overlayer and was further studied and applied to the case of
strained heteroepitaxy.\cite{Khor00,Korutcheva00,Prieto05,Xiang10} The
rearrangement of mono- to bilayer islands, of bilayer to three-layer islands
was established by Khor and Das Sarma by making use of Monte Carlo simulations
in the (1+1)D case. During the deposition the process of 2D-3D transformation
takes place in such a way that the bilayer islands are almost completely formed
at the expense of the atoms incorporated into the monolayer islands, most of
the atoms building the three-layers islands originate from the bilayer islands.
\cite{Khor00} As shown above, this mechanism of growth is expected to take place
in the case of ``stiff" materials and in particular in compressed overlayers. We
show that this behavior has the characteristics of a typical nucleation process
but only in compressed epilayers.

\begin{figure}[htb]
\includegraphics*[width=7.5cm]{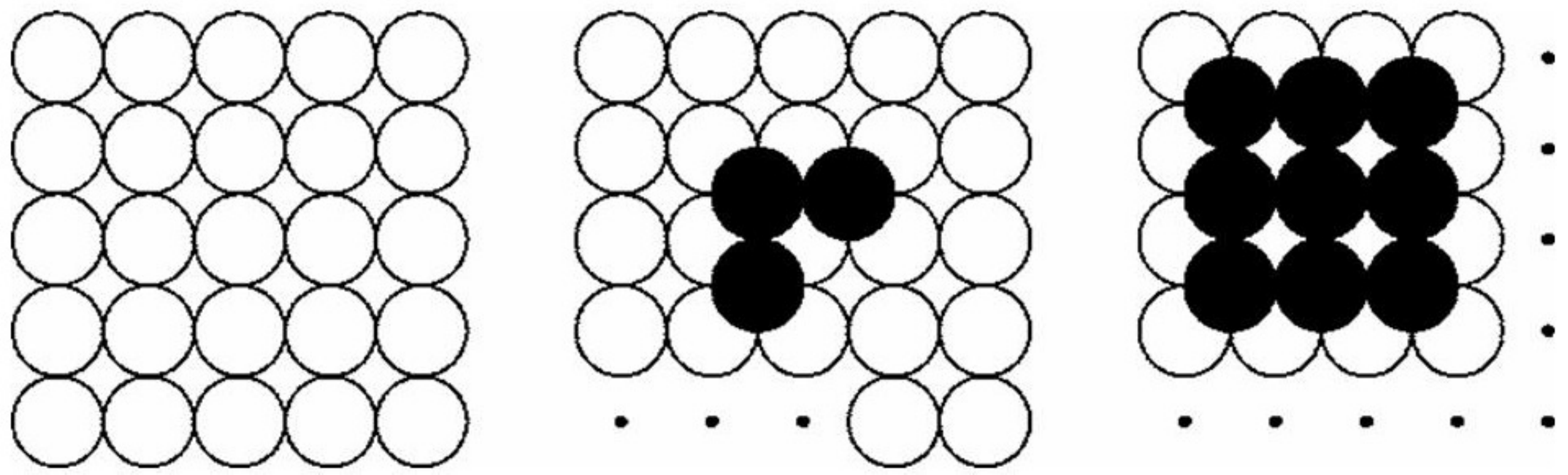}
\caption{\label{transform} Schematic representation of the process for the
evaluation of the activation energy of the monolayer-bilayer transformation. The
initial state is a square monolayer island. The intermediate state is a partial
bilayer island whereas the final state is a truncated bilayer pyramid. (J. E.
Prieto, I. Markov, Phys. Rev. B 72, 205412 (2005)). By permission of the
American Physical Society.}
\end{figure}

\begin{figure}[htb]
\includegraphics*[width=7.5cm]{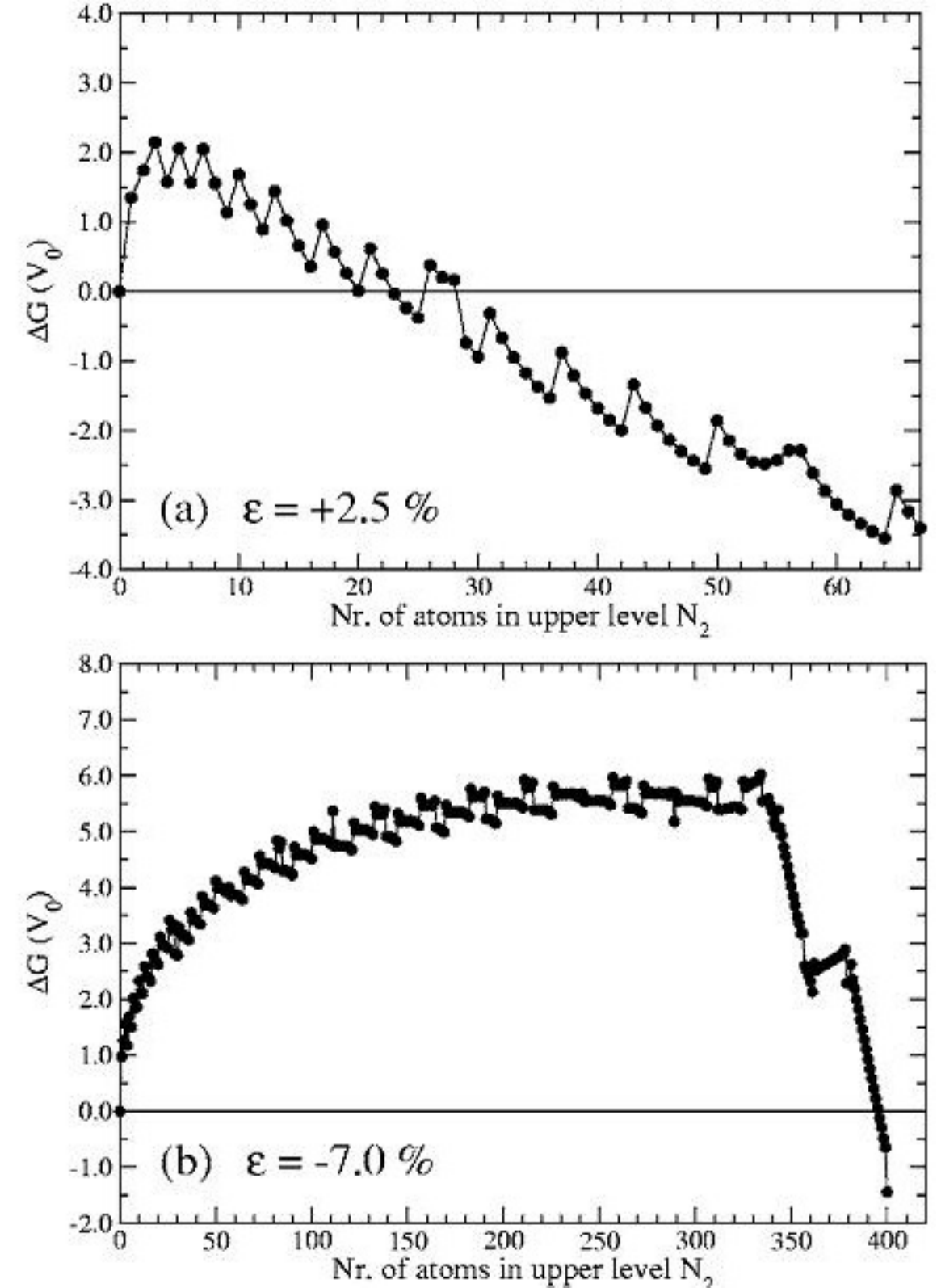}
\caption{\label{transcurves} Transformation curves showing the change
of energy in units of the bond energy $V_0$ as a function of the number
of atoms in the upper level for (a) positive (+2.5\%) and (b) negative (-7.0\%)
values of the misfit. The number of atoms in the initial monolayer island
$N_0=841=29 \times 29$ is chosen in a way to give a complete truncated bilayer
pyramid consisting of $21 \times 21=441$ atoms in the lower and $20 \times
20=400$ atoms in the upper level; $\mu =2\nu =36$ and the force constant $\gamma
= 648$. (J. E. Prieto, I. Markov, Phys. Rev. B 72, 205412 (2005)). By permission
of the American Physical Society.}
\end{figure}

For the study of the mechanism of mono-bilayer transformation, we simulate the
following imaginary process. We consider an initial monolayer island with a
square shape as shown in Fig.\ \ref{transform}. We detach atoms from its edges
and transfer them on top of the island thus building a compact second layer 
island at the center of the island underneath. The process proceeds 
until the second layer covers completely the first one thus producing a 
bilayer truncated pyramid. 
We calculate the the energy associated to the mono -- bilayer transformation 
by subtracting the energy of the initial monolayer island from the energy 
of the incomplete pyramid at every stage of the process.

\begin{figure}[htb]
\includegraphics*[width=7.5cm]{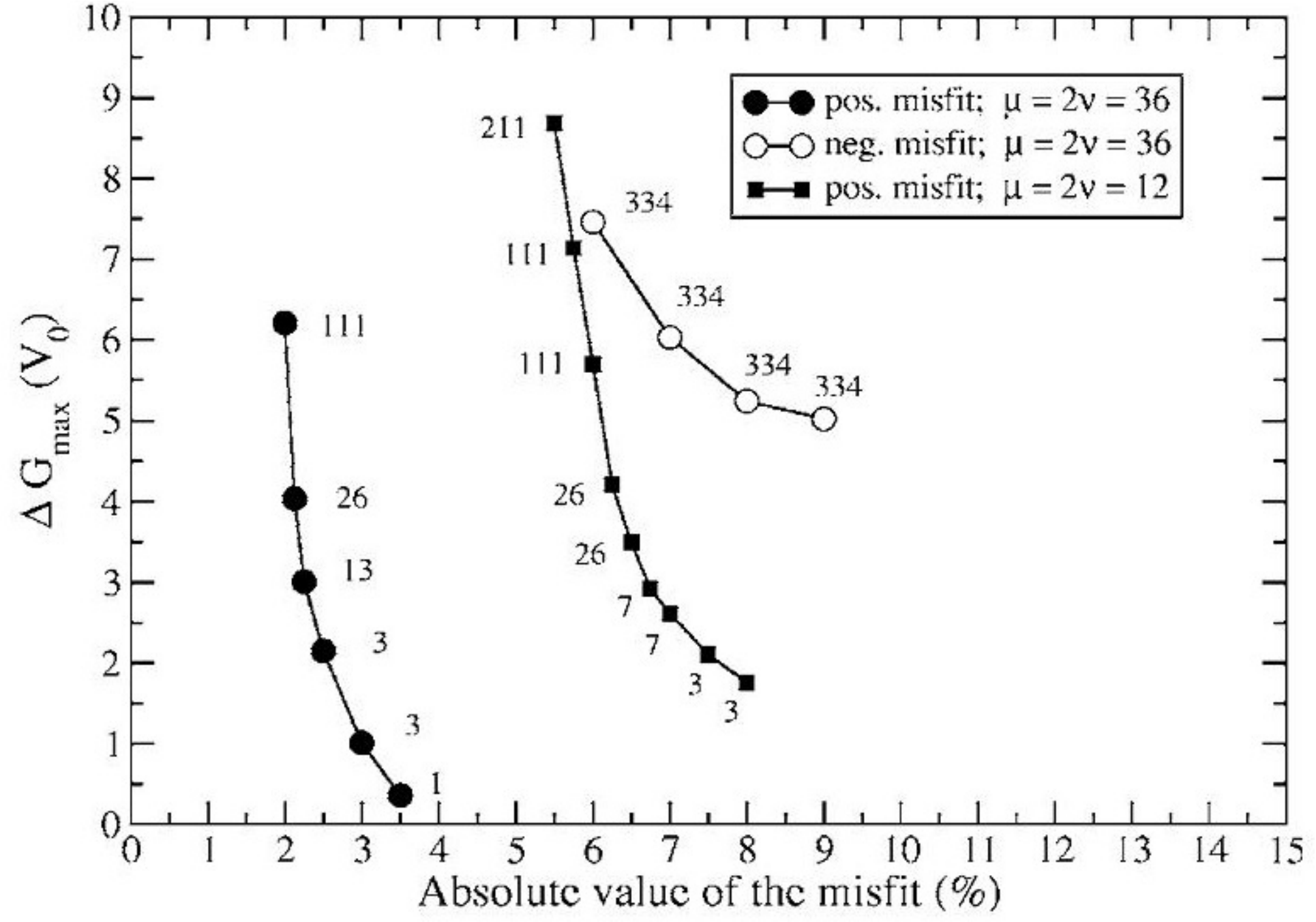}
\caption{\label{deltage} Height of the energetic barriers in units of $V_0$ as a
function of the lattice misfit in compressed and tensile overlayers. The figures
at each point show the number of atoms in the critical nucleus, $i^*$. The
initial island size was $29 \times 29 = 841$ atoms. (J. E. Prieto, I. Markov,
Phys. Rev. B 72, 205412 (2005)). By permission of the American Physical
Society.}
\end{figure}

Fig.\ \ref{transcurves} shows typical transformation curves of the energy
change associated with the transfer of atoms from the lower to the upper level
as a function of the number of atoms in the upper island for positive [Fig.\
\ref{transcurves}(a)] and negative [Fig.\ \ref{transcurves}(b)] values of the
misfit. In compressed overlayers, the transformation curve for $\Delta G$
exhibits the typical shape for the formation of the nucleus of a new phase. It
displays a maximum, $\Delta G_{\rm max}$, for a critical cluster size, $i^*$,
and then decreases beyond this size up to the completion of the transformation.
The atomistics of the transfer process (i.e. the completion of rows in the upper
level and their depletion in the lower one) are responsible for the
non-monotonic behavior of the curve.

Figure \ \ref{deltage} demonstrates the dependence of the height of the barrier
$\Delta G_{\rm max}$ on the misfit in compressed and tensile overlayers. The
figure at each point gives the number of atoms $i^*$. As seen $\Delta G_{\rm
max}$ decreases steeply with increasing misfit in a way similar to the decrease
of the work required for nucleus formation with increasing supersaturation in
the classical theory of nucleation.\cite{Markov17,Kashchiev00,Markov10} We can
accept a dependence of the form $\Delta G_{\rm max} = K f^{-n}$ where $K$ is a
constant proportional to the force constant $\gamma$, and $f$ is the lattice
misfit. Then we find $n = 4.29$ for $\mu = 2\nu = 12$ and $n = 4.75$ for $\mu =
2\nu = 36$. It is worth noting that considering 3D nucleation on top
of the wetting layer, Grabow and Gilmer predicted a value $n = 4$ for small
misfits (large nuclei) assuming $\Delta G_{\rm max}$ is inversely proportional
to the square of the supersaturation, which in turn is proportional to the
square of the lattice misfit.\cite{Grabow88} The same exponent of four was
obtained also by Tersoff and LeGoues.\cite{Tersoff94}

Looking at Fig.\ \ref{deltage} we note that the critical nuclei consist of a
number of atoms which exceeds by one atom the size of a compact cluster.
This is in accordance with the atomistic theory of nucleation where the
critical nucleus size is given by $i^* = i(i-1) +1$ ($i = 1, 2,
3\hdots$).\cite{Kashchiev08} The reason is easy to understand. The additional
atom creates two kink positions for the growth of the next row of atoms. Thus,
this additional atom can be considered as the one-dimensional nucleus which
gives rise to a new atomic row and thus transforms the rectangular island into a
square one. It is worth noting that one-dimensional nuclei cannot exist in the
thermodynamic sense but can be defined by making use of a kinetic
approach.\cite{Voronkov70,Frank74,Zhang90} For a recent review see
Ref.~(\onlinecite{Markov17})

In the case of tensile overlayers, the transformation basically increases 
all the way up to a number of atoms which is approximately equal to the 
number of atoms required to complete the upper level minus the number
of atoms necessary to build the last four edge rows of atoms. 
Thus the final collapse of the energy is due
to the disappearance of 8 single steps which repulse each other and the
formation of 4 low energy facets. The highest value of the number of atoms
before the collapse of the energy does not depend on the misfit, and as a
whole, the transformation curve displays a non-nucleation behavior. Having in
mind that the maximum number of atoms in tensile overlayers is higher than that
in compressed ones we can conclude that the process of 3D islanding will be 
very difficult as it will require (astronomically) very long times.

\begin{figure}[htb]
\includegraphics*[width=7.5cm]{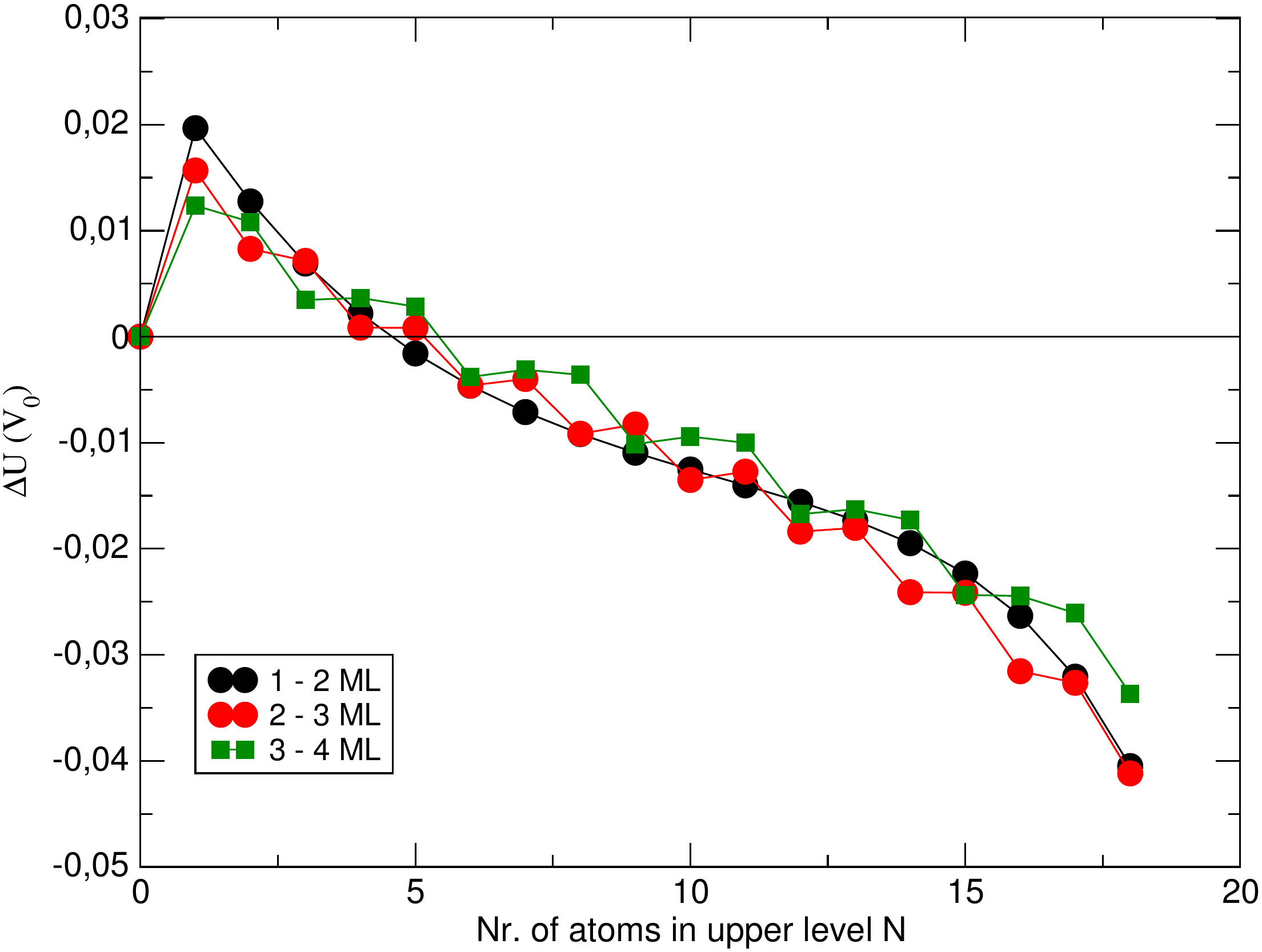}
\caption{\label{thick-islands} (Color on-line) Energy change $\Delta E_{\rm
n}$ in units of $V_0$ connected with the transformation of mono- to bilayer
islands (black circles), bi- to trilayer islands (red circles), and three- to
four-layer islands (black squares), as a function of the number of atoms $n$ in
the uppermost chain.}
\end{figure}

It is of interest to study the transformation to thicker islands. Figure.\
\ref{thick-islands} shows the energies of transformation of mono- to bilayer
islands, bilayer to three-layer island, and three- to four-layer islands. All
curves have the characteristic behavior of a nucleation process. As shown in
Ref. (\onlinecite{Stoyanov82}) the chemical potential of the upper island at the
maximum is exactly equal to that of the initial monolayer island, and the
supersaturation with which the nucleus of the second layer is in equilibrium is
equal to the difference of the energies of desorption of the atoms from the like
and the unlike substrate. This is, namely, the
driving force for the 2D-3D transformation to occur. Note that the (1+1)D model
is in fact one dimensional and the nuclei do not exist in the thermodynamic
sense because the length of a row of atoms does not depend on the
supersaturation as discussed above.\cite{Voronkov70,Frank74,Zhang90} However,
considering our (1+1)D model as a cross section of the real (2+1)D case, we can
treat the curves in Fig.\ \ref{thick-islands} as dependences of the free
energy for nucleus formation and growth or, as consecutive transformation
curves. We would like to emphasize that in the (2+1)D model the nucleus does not
necessarily consist of one atom. Its size must depend on the lattice misfit,
and in a real situation on the temperature. The curves describing the 2-3 and
3-4 transformations behave in the same way but the work for nucleus formation
(the respective maxima) decrease with the thickening of the islands.
This means that the mono-bilayer transformation is the rate-determining
process for the total mono-multilayer (2D-3D) transformation.

\section{Multilayer growth of 3D islands}

\begin{figure}[htb]
\includegraphics*[width=7.5cm]{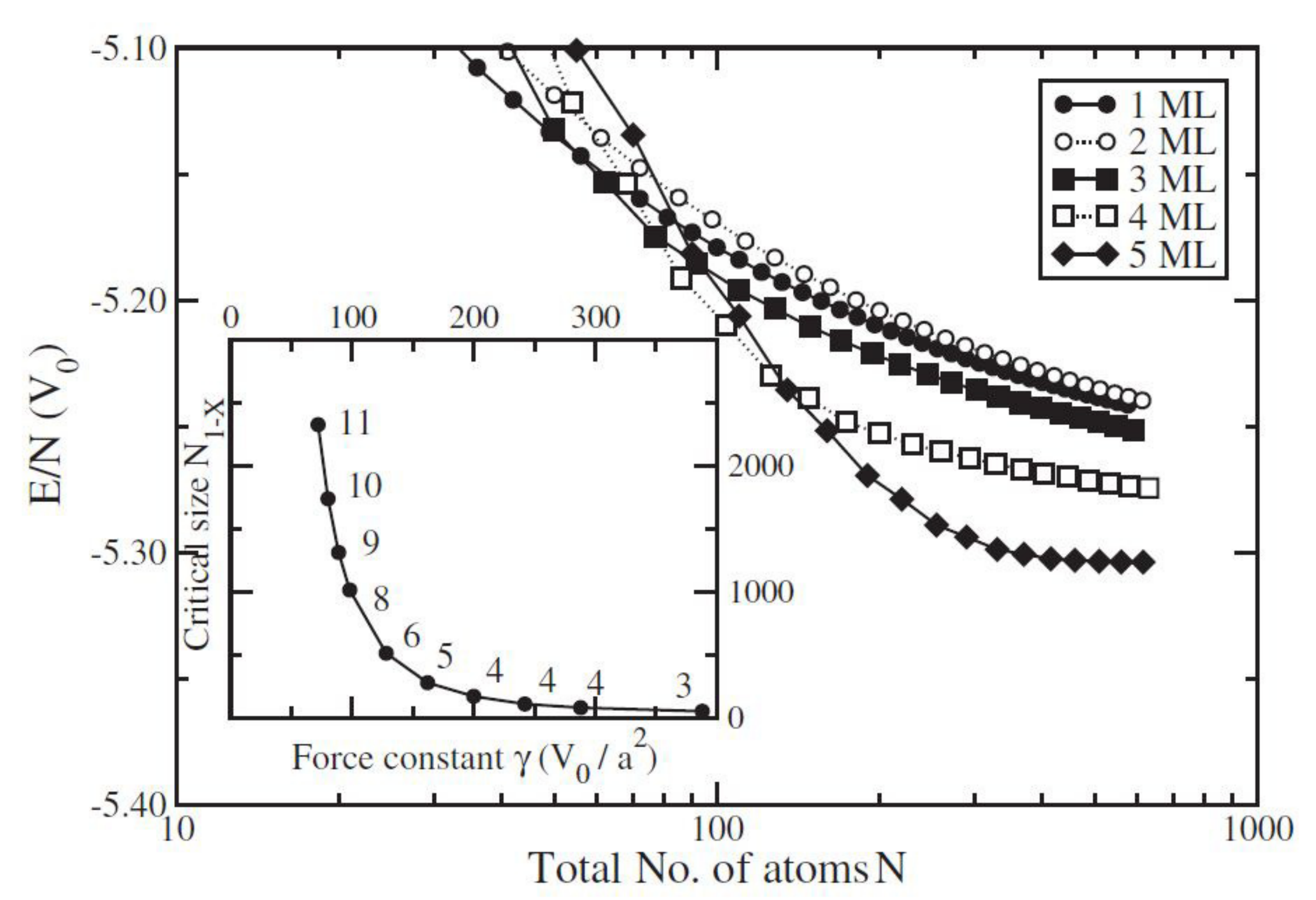}
\caption{\label{monomulti} Energy per atom of mono- and multilayer tensile
islands as a function of the total number of atoms. The misfit amounts to -7\%
and $\mu =2\nu = 26$. The inset shows the dependence of the critical size
$N_{\rm 1-X}$ on the force constant $\gamma$. $N_{\rm 1-X}$ is the size beyond
which 1-ML islands become unstable against X-ML islands. The thickness X is
denoted by the numbers at each point. (J. E. Prieto, I. Markov, Phys. Rev. Lett.
98, 176101 (2007)). By permission of the American Physical Society.}
\end{figure}

As mentioned in Section V in the case of negative misfits (tensile overlayers)
a decrease of the force constant $\gamma$ leads to the consecutive disappearance
of the crossing points $N_{\rm 12}, N_{\rm 13}, N_{\rm 14}\hdots$. As observed
in Fig.\ \ref{monomulti} the reason is that the energy per atom of multilayer
islands (in the particular case shown, of bilayer islands) remains always larger
than that of monolayer islands.\cite{Prieto07} The energy curves never cross, 
irrespective of the islands size. The energy curve for monolayer islands is
crossed by the curve of the energy of the three-layer islands at a size $N_{\rm
13}$. The reason for the energy of the bilayer islands to be always larger than
that of the monolayer islands is the weaker strain relaxation at the double step
edge of the laterally smaller island of the same total number of atoms. The
island must become of triple height in order for the strain relaxation to
prevail over the step energy. This is in accordance with the considerations of
van der Merwe {\it et al}.\cite{Merwe86} Thicker island consisting of $n$ layers
have an effective force constant, $\gamma_{\rm n}$, larger than the force
constant, $\gamma$, of a single monolayer. Decreasing the material's stiffness
(decreasing $\gamma$) leads to the necessity of increasing the threshold
thickness or, in other words, the effective force constant $\gamma_{\rm n}$.
This is shown in the inset in Fig.\ \ref{monomulti} which gives the dependence
of the critical size $N_{\rm X-1}$ (beyond which single ML islands become
unstable against X-ML islands) on the force constant $\gamma$. The critical
thickness below which multilayer islands are energetically unfavored sharply
increases, together with the critical volume, for decreasing force constant
$\gamma$.

\begin{figure}[htb]
\includegraphics*[width=7.5cm]{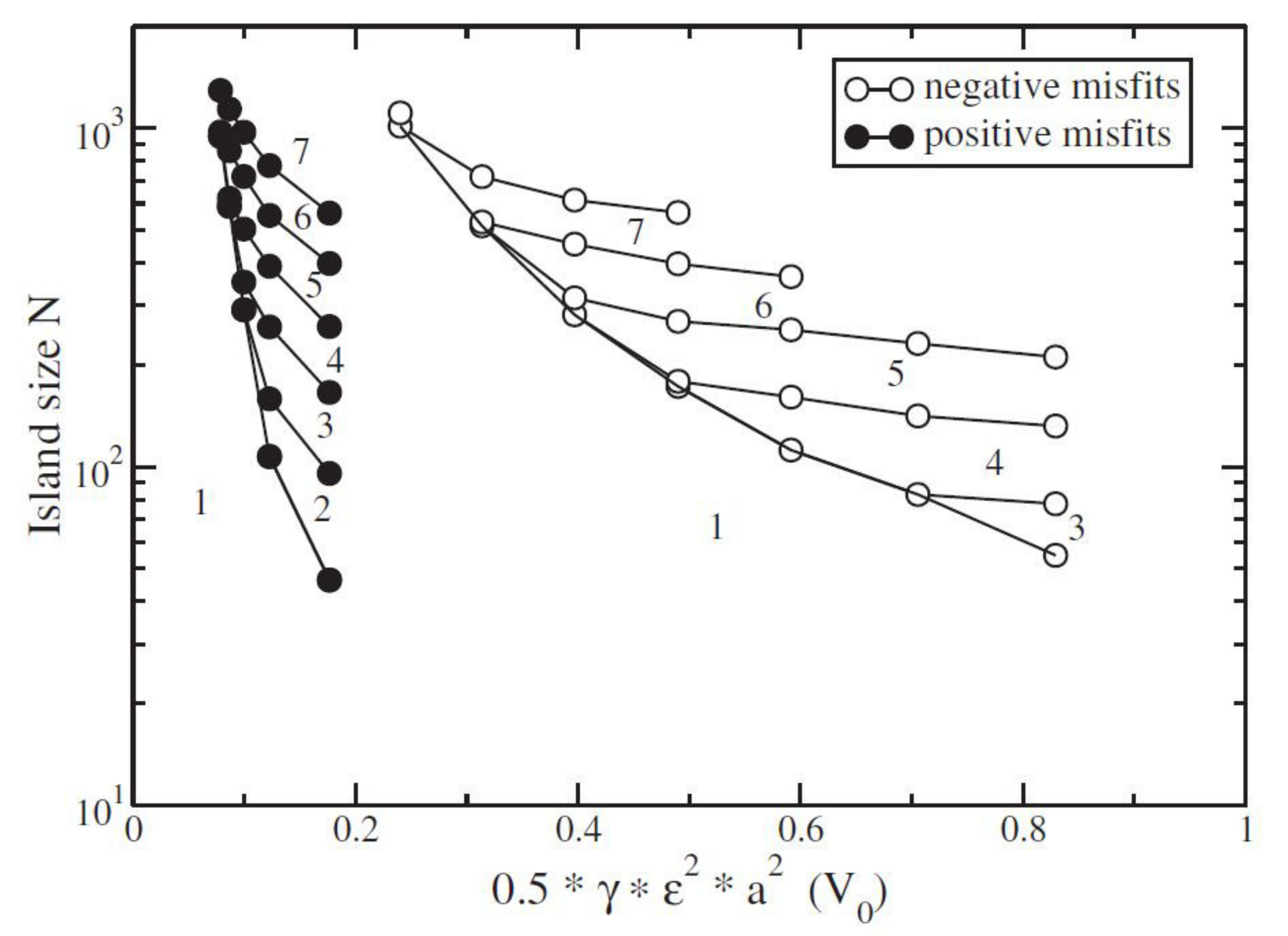}
\caption{\label{phasedia} Phase diagram in coordinates total number of 
atoms $N$ vs. bulk strain energy per bond ${\cal E} = 0.5\gamma f^2 a^2$, 
for positive and negative values of the misfit and for potentials with 
$"\mu = 2\nu$. The numbers mark the regions of stability of islands of the
corresponding number of monolayers, regions which are limited by the displayed 
curves. [J.E. Prieto, I. Markov, Phys. Rev. Lett. 98, 176101 (2007)]. 
By permission of the American Physical Society.}
\end{figure}

This is illustrated in Fig.\ \ref{phasedia}. It shows a phase diagram
of stability of mono- and multilayer islands in coordinates of island size $N$
vs. strain energy per bond ${\cal E} = 0.5\gamma f^2 a^2$. The numbers in 
the plot give the heights of the stable islands inside the corresponding 
regions, limited by the curves shown. Single monolayer
islands are stable at small numbers of atoms. Bilayer islands are stable at
large strain energies i.e. at large force constant or large misfits. Otherwise,
thicker islands will become stable. Islands thinner than a certain number
of layers will be forbidden for thermodynamic reasons. Tensile overlayers
require larger absolute values of the misfit compared with compressed ones
in order to compensate the weaker attractive forces between the atoms.

\begin{figure}[htb]
\includegraphics*[width=7.5cm]{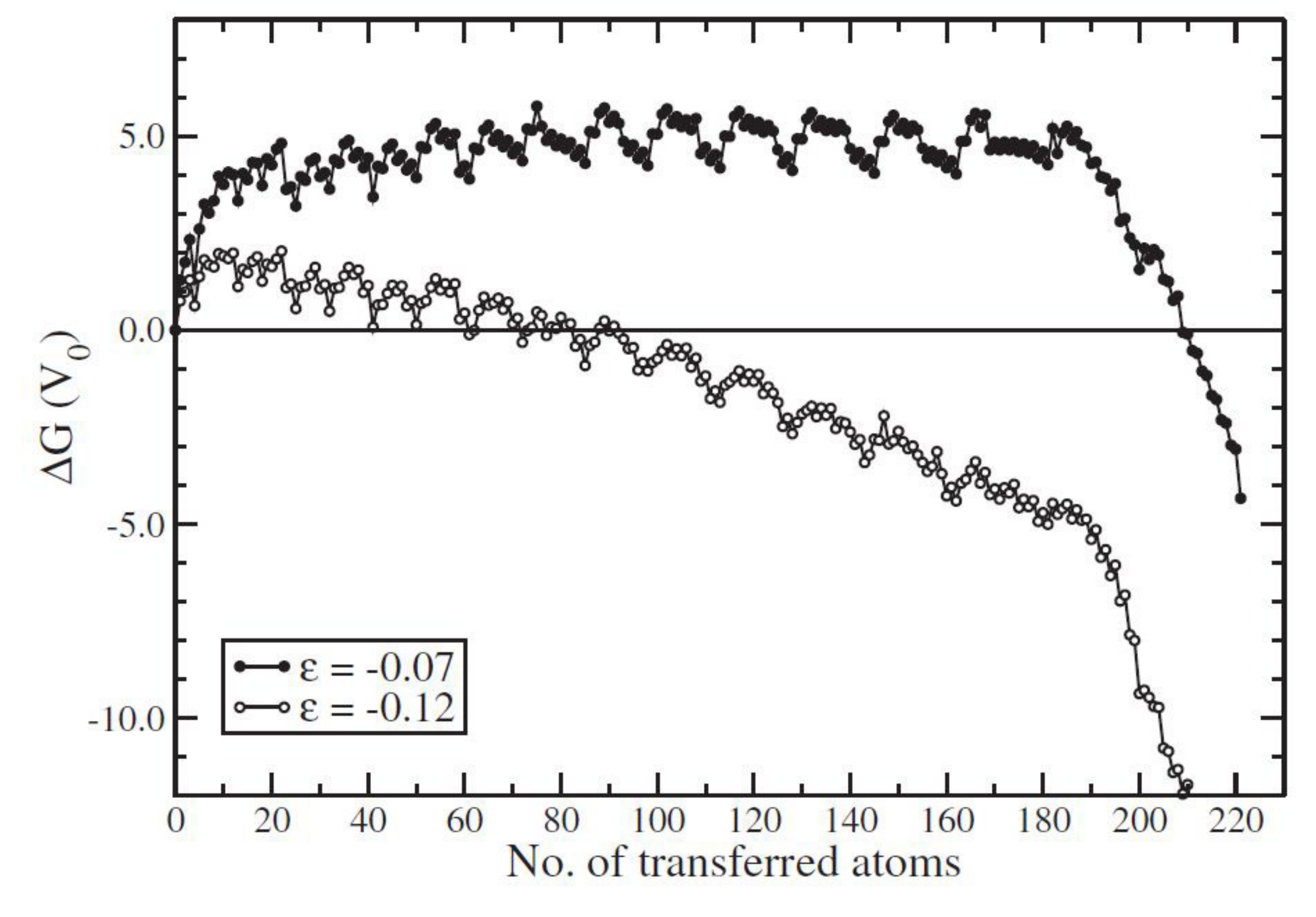}
\caption{\label{multigrowth} Curves showing the energy change of transformation
from 1 to 3 layers high islands in units of the bond energy $V_0$ as a function
of the number of atoms transferred to the upper level. Upper curve: $f = - 7\%$,
lower curve: $f = - 12\%$. The number of atoms in the initial monolayer island
($365 \approx 19 \times 19$) gives a complete truncated three-layer pyramid ($12
\times 12 + 11 \times 11 + 10 \times 10$) and $\mu = 2\nu = 26$. (J. E. Prieto,
I. Markov, Phys. Rev. Lett. 98, 176101 (2007)). By permission of the American
Physical Society.}
\end{figure}

The results shown above suggest a novel mechanism of transformation of
monolayer to multilayer islands which differs from the ordinary layer-by-layer
growth described above. An X-layer island will form by the formation and 
lateral, two-dimensional growth of an (X-1)-layer island. 
Figure\ \ref{multigrowth} shows
the transformation curves of mono- to three-layer islands by formation and
lateral growth of bilayer island. Two curves are shown, for misfits of -7\% and
-12\%. Atoms are detached from the edges of the initial monolayer island and are
incorporated into the double steps of the bilayer island growing on top. The
low-misfit curve is similar to the layer-by-layer curve shown in Fig.\
\ref{transcurves}(b). The energy tends to increase all the way and shows at the
very end a sudden collapse due to the disappearance of the single and double
steps to produce low-energy facets. In the case of larger misfit (larger
strain energy per bond) the curve shows a nucleation behavior. The bilayer
nucleus $i^*$ consists of 22 atoms.

In order to better understand the multilayer mode of 2D-3D transformation, let
us consider it in more detail. We consider for simplicity the mono-bilayer
transformations. The physics of the transformation of mono- to
multilayer island is essentially the same. During the process of transformation
the overall step length increases; the step length of the lower island
decreases and that of the upper island increases in such a way that the total
step length increases. This can be easily shown assuming square or circular
islands. The variation of the step length is very important since strain is
relaxed at the steps. Another effect associated with the increase of the
step length is the increase of the step energy since the number of the dangling
bonds increases. In addition, the vertical displacements of the atoms at the 
steps lead to an increase of the wetting parameter, in other words, of the 
tendency to 3D islanding. Still another effect is the step repulsion energy, 
which increases as $l^{-2}$, where $l$ is the step separation.\cite{Nozieres92}
This plays a significant role only near the end of the transformation. The
strain relaxation and the vertical displacements of the atoms close to the step
edges favor the process of 3D islanding, i.e. the mono-bilayer
transformation.\cite{Prieto05,Muller96} The increase of the step energy and the
step-step repulsion oppose it. Then, in order 3D islands to form it is necessary
the first two factors to prevail. If the misfit is negative and the film
material is soft (small value of the force constant $\gamma$) the second layer
nucleus must be thicker in order to give rise to an effectively larger value of
$\gamma$ and in turn to greater vertical displacements.

\begin{figure}[htb]
\includegraphics*[width=7.5cm]{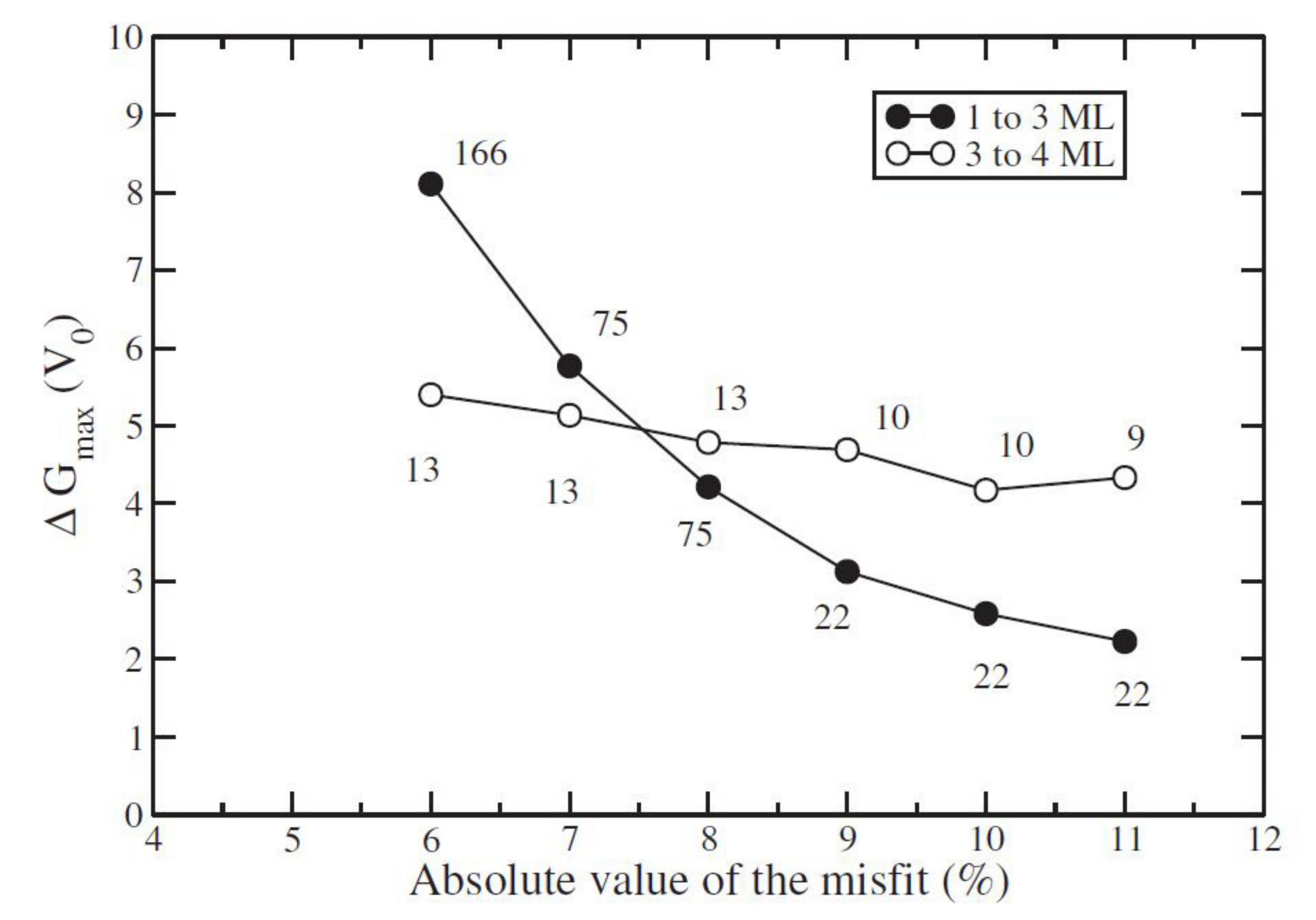}
\caption{\label{multimono} Nucleation barriers in units of $V_0$ as a function
of the absolute value of the negative misfit, for transformation from
single monolayer to three-layer islands. The number of atoms in the critical
nucleus is given at each point. The barrier for monolayer nucleation on top of
the three-layer island is also shown. The total number of atoms is 365, and $\mu
= 2\nu = 26$. (J. E. Prieto, I. Markov, Phys. Rev. Lett. 98, 176101 (2007)). By
permission of the American Physical Society.}
\end{figure}

Once a pyramid of height X MLs is formed, it can continue to grow further in
height by the formation and growth of monolayer nuclei. It is thus of interest 
to
study the formation of 2D nuclei on the upper surface of the pyramid. Figure\
\ref{multimono} shows a comparison of the nucleation barrier of mono-three-layer
(1-3) transformation with the barrier for monolayer nucleation on top of the
three-layer pyramid (3-4 transformation). The monotonic decrease of the
nucleation barrier with misfit for the 1-3-ML transformation is characteristic
of nucleation behavior.\cite{Prieto05} The barrier for the 3-4-ML transformation
is smaller than for 1-3-ML at small misfits. This means that if a 1-3
transformation takes place, a forth layer can be formed on top. On the contrary,
at larger absolute values of the misfit, the barrier for formation of a nucleus
on the top surface of the three-layer pyramid is significantly greater than the
barrier for formation of a bilayer nucleus on the first layer, so that the
growth of the 4th atomic level might be strongly inhibited for kinetic reasons.
If this is the case the multilayer pyramid will grow laterally keeping its
height constant. Adatoms from the surface will prefer to join the side walls
instead of nuleate on the smooth top surface. This is more important for 
tensile islands than for compressed ones.

\section{Comparison with experiment}

In this section we verify two predictions of the model, the existence of a
critical misfit below which coherent 3D islands cannot be formed on the stable
wetting layer, and the existence of multilayer transformation of monolayer into
3D islands.

\subsection{Critical misfit for coherent Stranski-Krastanov growth}

As shown in Figs.\ \ref{en12} and \ref{en123D}, coherent 3D islands can be
formed on the stable wetting layer only at misfits larger than a critical
misfit $f_{\rm cr}$ particularly in compressed overlayers. In the important
cases of Ge/Si\cite{Voigt01,Eagles95} and InAs/GaAs\cite{Joyce04,Wu15}, the
overlayer is compressed by 4\% and 7\%, respectively. In fact, the
critical misfit for mono-bilayer transformation and its dependence on misfit
sign was first suggested on the base of computer simulations on a (1+1)D 
model by Korutcheva {\it et al.}\cite{Korutcheva00}

Xie {\it et al.}\cite{Xie94} found a critical compressive misfit of 1.4\% for 3D
islanding upon deposition of Si$_{0.5}$Ge$_{0.5}$ films on relaxed buffer layers
of Si$_{x}$Ge$_{1-x}$ varying $x$ from 0 (pure Ge) to 1 (pure Si). In this way
they covered the whole range of misfit from -2\% to 2\%. Films under tensile
misfit were found always stable against 3D islanding.

Pinczolits {\it et al.}\cite{Pinczolits98} have found that upon deposition of
PbSe$_{1-x}$Te$_{x}$ on PbTe(111) the overlayer remains purely two-dimensional
when the misfit is less than 1.6\% in absolute value (Se content $<$ 30\%).
Note that contrary to the systems Ge/Si and InAs/GaAs, the lattice misfit of the
system PbSe/PbTe is tensile (-5.3\%). One should have in mind, however, that
PbSe could be considered as a ``stiff" material.\cite{Petersen14}

Leonard {\it et al.} \cite{Leonard93} have successfully grown coherent 3D
islands of In$_{x}$Ga$_{1-x}$As on GaAs(001) with $x = 0.5$, or $f \approx
3.6\%$ but 60 {\AA} thick 2D smooth planar films at $x = 0.17$ ($f \approx
1.2\%$). Thus a critical misfit between 1.2\% and 3.6\% should exist.

The above result has been confirmed by Walther {\it et al.}\cite{Walther01}
who found that a critical misfit of about 1.8\% has to be exceeded in order 3D
islands to grow in the Stranski-Krastanov mode, or a critical In content of 
approximately $x = 0.25$ to be exceeded in order In$_{x}$Ga$_{1-x}$As
quantum dots to grow on top of the wetting layer on GaAs(001).

It is interesting to note that the critical misfit for 3D islanding to take
place covers a great part of the maximal misfit of the respective system. Thus
it varies from 0.25\% for InAs/GaAs (1.8\% from 7.2\%) to 0.33\% for Ge/Si
(1.4\% from 4.2\%). The latter means that the layer-by-layer growth cannot be
considered as a rare event as was thought before. This behavior of the
systems studied and in particular the existence of a critical misfit cannot be
explained either by the nucleation theory of 3D islands formation or by the
barrierless concept.

\subsection{Forbidden island heights in strained heteroepitaxy}

The last two decades have witnessed intensive studies of epitaxial growth of
metals in particular Pb, Ag and Al, on semiconductors surfaces at low
temperatures (130 - 180 K).\cite{Gavioli99,Hupalo01,Su01,Floreano03} Flattop Pb
islands with a preferred height of 7 monolayers were observed to grow on the
wetting layer on Si(111)$7\times 7$.\cite{Gavioli99,Hupalo01,Su01,Yeh00} These
observations were explained in terms of the energy decrease owing to electron
confinement and spilling of charge over the metal-semiconductor interface or a
``quantum size effect" (QSE) by Zhang {\it et al.}, who coined for this reason
the term ``electronic growth".\cite{Zhang98} The thickness of islands was found 
to be in the
range of four to nine atomic layers; among these islands, those with a height
of seven ML were clearly observed to dominate; Pb islands on Si(111) with
thicknesses ranging from 1 to 3 MLs were never
observed.\cite{Su01,Chang02,Ozer05} Flattop islands with a preferred height grow
laterally without thickening.\cite{Gavioli99,Ozer05,Budde00} It was also
observed that 2-ML thick flattop Ag islands on Si(111) increase linearly in
size preserving their height,\cite{Gavioli99} whereas single layer islands
preserve a nearly constant size of 500\AA.\cite{Su05}

The above observations can be explained by the quantum size effect, but we show
that classical effects associated with strain relaxation at steps and the
interplay of strain and edge energies can give a plausible explanation as
well.\cite{Prieto07}

Obviously, the first thing to do is to estimate the strain energy per bond
${\cal E}$ for the metals under study in the experiments Pb, Ag and Al. For this
aim we need an estimate of the respective force constant by making use of the
relation $\gamma = Eb/2(1+\mu_{\rm P})$ where $E$ and $\mu_{\rm P}$ are the
Young modulus and the Poisson ratio of the overlayer material. We get the values
0.15, 3.08 and 2.27 in units of $V_0$ for Pb, Ag and Al, respectively. With a 
value of 0.15 for Pb we expect a transition from 1 to 8-9 MLs in reasonable
agreement with the experimental value of preferred height of 7
MLs.\cite{Gavioli99,Hupalo01,Su01,Yeh00,Chang02,Ozer05,Su05}

The very high values of ${\cal E}$ estimated for Ag and Al are due to the 
expected failure of the harmonic approximation for high values of the misfit.
Anyway the prediction is the presence of thinner islands, as observed in
experiment.\cite{Su05,Liu04} Note that our estimations are performed on coherent
islands with (100) orientation, while most experiments are performed on Si(111).
However, our estimations should be approximately valid, because the relaxation
of the strain energy must be partially balanced by disregistry.

The lateral growth of 2-ML high Ag islands until coalescence\cite{Gavioli99} can
be explained by inhibited nucleation of 2D islands on top of the bilayer
islands. The constant size of 500$\AA$  of the monolayer islands observed
experimentally\cite{Su05} can be considered as the critical size $N_{\rm
12}$. Clear evidence for a transformation process is the rearrangement of 2- to
3-ML high Fe islands on Cu$_3$Au(001) deposited at 140 K when annealed at 400
K.\cite{Canepa00,Verdini02}

Finally, based on the above considerations, we predict that soft tensile metal
overlayers should undergo a multilayer transformation whereas stiffer compressed
metals overlayers should show a layer-by-layer transformation. Thus we can
expect that In with ${\cal E} = 0.17V_0$ should undergo a multilayer
transformation similar to that of Pb/Si(111) but with a slightly smaller 
preferred height. Indeed, Chen {\it et al.} observed the formation of 
flattop In islands with a preferred height of 4 MLs.\cite{Chen08}

\section{Discussion}

As follows from the discussion above, Bauer's criterion of wettability
of the substrate by the overlayer material can be successfully applied 
to the Stranski-Krastanov growth mode. 
We will show that the idea of the trade-off of the strain
relaxation energy and the surface energy does not contradict the concepts
developed in this paper. We do not discuss the barrierless transformation of
ripples as one possible additional mechanism of formation of 3D islands.

The trade-off between strain and surface energies is a macroscopic concept.
The development of an effective positive wetting parameter owing to the lattice
misfit is a microscopic phenomenon. If we assume that the stable wetting layer
grows in a layer-by-layer mode by the formation and lateral spreading of 
2D nuclei,
it is logical to suppose that further deposition will proceed in the same way.
The resulting 2D islands are enclosed by step edges. Strain relaxation takes
place at these steps. Atoms at the edges are displaced from their positions in
the bottoms of the corresponding potential troughs and the atomic separations at
the edges and corners of the islands are very close to the natural separation 
of the bulk deposit crystal. 
For this reason the nucleation of the second layer takes
often place at the edges and corners as ``$\hdots$atoms are happy to be there,
because they find an atomic distance they would like to
have",\cite{Villain05}(see also Ref. (\onlinecite{Prieto11})). Experimental
evidence supports this conclusion.\cite{Shi06,Murray95,Molina07}

The smaller the misfit, the smaller the displacements of the edge atoms and
in turn, the stronger is the average wetting (the smaller is the effective
wetting parameter). This leads to the appearance of a critical misfit
below which the wetting parameter is too small and the energy of the bilayer
islands is always larger than that of the monolayer ones (see Fig.\
\ref{smallmisfit}). The curves of the energies of mono- and bilayer islands
do not cross each other. The formation of coherent 3D islands becomes
thermodynamically unfavored. The film will continue to grow in a 2D mode until
the strain is relaxed by the introduction of misfit dislocations. In order for
the mono- and bilayer islands energies to cross each other to give rise to 3D
islanding, the misfit and in turn the strain relaxation at the edges must 
be larger.

\begin{figure}[htb]
\includegraphics*[width=7.5cm]{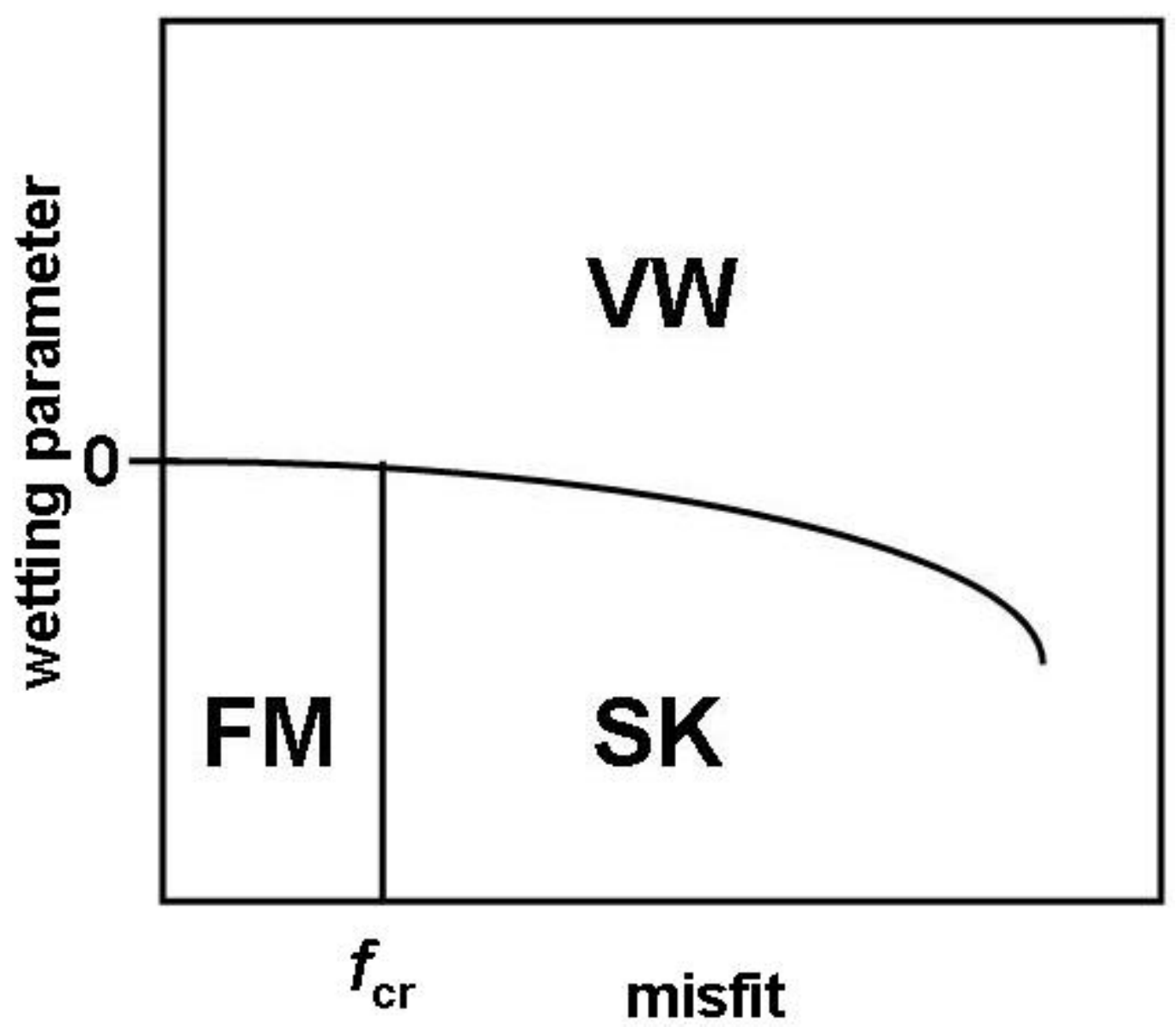}
\caption{\label{phase_diagram} Schematic representation of the phase diagram
for the occurrence of the VW, FM and SK growth modes.}
\end{figure}

We conclude from the above consideration that the critical misfit, $f_{\rm
cr}$, represents in fact the dividing line between the FM and SK growth modes. 
We then can devise a phase diagram for the occurrence of any mode of growth in
coordinates wetting parameter - lattice misfit (see Fig.\ \ref{phase_diagram}).
Volmer-Weber mode of growth takes place at positive values of the wetting
parameter. The curve slightly decreases with increasing misfit thus
widening the field for VW growth because of the contribution of the misfit to
the wetting as discussed above. Both FM and SK modes take place below this line,
FM at misfits smaller than the critical misfit and the SK mode at larger 
values of the misfit. 
Note that in tensile overlayers the dividing line between the FM
and SK modes should be placed at greater absolute values of the misfit. This
phase diagram differs from the one suggested by Grabow and Gilmer\cite{Grabow88}
by the displacement of the FM growth from zero misfit to the critical misfit,
$f_{\rm cr}$, as discussed above. Note that in our case $f_{\rm cr}$ can be
larger or smaller depending on the misfit sign, and more importantly, on the
material stiffness. In the case of very soft materials and negative misfits,
$f_{\rm cr}$ can be so large that Stranski-Krastanov growth may not take place
at all. It is noteworthy that the critical misfit in alloy films could reach 
a significant value of about 30\% of the misfit between the pure binaries 
(Ge/Si or InAs/GaAs).

The average adhesion (the wetting) depends strongly on the anharmonicity of the
interatomic forces. Tensile islands adhere more strongly to the wetting layer
and the critical misfit beyond which coherent 3D islanding is possible is much
larger in absolute than in compressed overlayers. As a result, coherent 
SK growth in tensile films could be expected at very (sometimes
unrealistically) large absolute values of the negative misfit. The critical
misfit, however, depends on the material parameters (degree of anharmonicity,
strength of the chemical bonds, etc.) of the particular system and the 
formation of coherent quantum dots in tensile overlayers cannot be completely 
ruled out.

The weaker average adhesion in compressed overlayers leads to another effect at
misfits larger than the critical one. Owing to the stronger interatomic
repulsive forces, the edge atoms in compressed monolayer islands adhere 
weaker to the wetting layer compared to expanded islands.
This results in an easier transformation of mono- to bilayer islands, which is
the first step to the complete 2D-3D transformation. The latter includes also
kinetics in the sense that the edge atoms have to detach and form the upper
layers. However, it is not the strain at the edges (which is nearly zero) that
is responsible for the easier detachment of the edge atoms as suggested by
Kandel and Kaxiras\cite{Kandel95} but the weaker adhesion. The 2D-3D
transformation is hindered in tensile overlayers as the edge atoms adhere more
strongly to the wetting layer.

On the other hand, the existence of such critical sizes, which determine the
intervals of stability of islands with different thicknesses, could be
considered as the thermodynamic reason for the narrow size distribution of 3D
islands which is observed in experiments. This does not mean that this is the
only reason. Elastic interactions between islands and growth kinetics can have
stronger effects than thermodynamics. The 2D-3D transformation takes place
by consecutive nucleation events, each one needing to overcome a lower energetic
barrier than the preceding one. Thus, the mono-bilayer transformation appears as
the rate-determining step. This is easy to understand. Thicker islands have
larger values of the wetting function which facilitates the transformation from
$X$ to $X+1$ MLs high islands.

We conclude that the criterion of Bauer describes well the transition from
planar growth to 3D islanding in the Stranski-Krastanov growth mode as a
transition from the FM to the VW growth mode. 
The only difference from the classical VW
growth is that in the latter the adhesion parameter is constant and is due to
difference in chemical bonding. In the case of the coherent SK mode the chemical
bonding is the same and the nonzero adhesion parameter is due to the 
misfit and depends on the island thickness.

Let us now try to answer the question asked in the Introduction. As discussed
above, the driving force for the 3D island formation is the relaxation of
elastic stress: the island nucleates because the elastic energy per atom in the
wetting layer is larger than in the island. The relaxation of the strain in 3D
islands with respect to the wetting layer overcompensates the surface energy of
the side facets of the 3D crystallites. This is the main idea of the nucleation
concept. However, the elastic stress in the initial 2D islands on top of the
wetting layer gives rise to a positive effective wetting parameter. The latter
is the thermodynamic driving force for 3D islanding in the sense of the
wettability criterion of Bauer. We conclude that the 3D islanding in SK growth
mode obeys the criterion of Bauer owing to the particular structure of the
interface between the stable wetting layer and the 2D islands on top which
appear as precursors of the 3D crystallites.\cite{Frank49,Frank491}

In summary, we conclude that the Stranski-Krastanov growth mode appears as a
sequence of Frank-van der Merwe and Volmer-Weber growth modes. The wetting layer
and the 3D islands represent different phases in the sense of Gibbs which cannot
be in equilibrium with each other. The separate monolayers which belong to the
wetting layer represent also different phases which have different equilibrium
vapor pressures. Monolayer-high islands with a critical size appear as necessary
precursors for 3D islands. The 2D-3D transition takes place through a series of
intermediate states with discretely increasing thickness that are stable in
separate intervals of volume in the case of ``stiff" materials or by the
formation and lateral growth of 2D multilayer nuclei in the case of ``soft"
materials like Pb and In. At sufficiently large misfits, the barrier for 2D
multilayer nucleation is significantly smaller than the barrier for the
subsequent single-layer nucleation. Then islands with a preferred height will
continue to grow laterally instead of growing in height. 
There exists a critical misfit
below which coherent 3D islands are thermodynamically unfavored and the misfit
is accommodated by misfit dislocations at a later stage of growth. Coherent 3D
islands can only form at misfits larger in absolute value than the critical
misfit. Compressed overlayers show a greater tendency to 3D clustering than
expanded ones, in agreement with experimental results. The mechanism of the
layer-by-layer transformation in compressed overlayers is nucleation-like due to
the interplay of relaxation of the in-plane strain, which is proportional to the
total edge length and the increase of the total edge energy and the repulsion
between the edges. The critical nucleus consists of one atom in addition of a
compact shape; it plays the role of a one-dimensional nucleus giving rise to a
new atomic row. The compact shape is in general a rectangle with edges of $i$
and $i - 1$ atoms, while square shapes can also appear if the length of the
critical nucleus is comparable to the number of atoms in the edge of the
original first-layer island. In some cases nuclei in the upper layer can form
on the edges or the corners of the underlying 2D island as the atom separations
there are nearly the same as in the bulk deposit crystal. The transformation
curve in tensile overlayers shows a ``non-nucleation" behavior characterized by
an overall increase of the energy up to the stage when the single steps coalesce
to produce low-energy facets. This is accompanied by a collapse of the energy.

This research did not receive any specific grant from funding agencies in the
public, commercial or not-for-profit sectors.

\end{document}